\documentclass[prd,aps,
longbibliography,
nofootinbib,
floatfix,
superscriptaddress]{revtex4}
\usepackage{graphicx}
\usepackage{epsfig}
\usepackage{rotating}
\usepackage{amssymb}
\usepackage{dsfont}
\usepackage{psfrag}
\usepackage{amsmath,euscript,array,mathrsfs}
\usepackage{bbold}
\usepackage{bm}
\usepackage{multirow}
\usepackage{dcolumn}
\RequirePackage{doi}
\usepackage[skip=10pt plus1pt, indent=20pt]{parskip}
\usepackage{float}
\usepackage{cancel}
\floatstyle{plaintop}
\restylefloat{table}
\usepackage{bbding}
\usepackage{orcidlink}
\usepackage{hyperref}
\usepackage{tikz}

\usepackage{xcolor}
\colorlet{RED}{red} 

\newcommand{\orcidauthorPIAI}{0000-0002-2251-0111} 
\newcommand{\orcidauthorFATEMIABHARI}{0000-0003-1369-6505}

\newcommand{\beq}{\begin{equation}}
\newcommand{\eeq}{\end{equation}}
\newcommand{\beqs}{\begin{eqnarray}}
\newcommand{\eeqs}{\end{eqnarray}}
\newcommand{\lsim}{\mathrel{\raisebox{-
.6ex}{$\stackrel{\textstyle<}{\sim}$}}}

\renewcommand{\L}{{\cal L}}

\newcommand{\A}{{\cal A}}

\def\hbar{\hspace{0pt}\raisebox{1pt}{$-$} \hspace{-7pt} h}

\def\di{\mbox{d}}
\def\r{\rho}

\newcommand{\be}{\begin{equation}}
\newcommand{\ee}{\end{equation}}

\newcommand{\bea}{\begin{eqnarray}}
\newcommand{\eea}{\end{eqnarray}}

\def\lbldef#1#2{\expandafter\gdef\csname #1\endcsname {#2}}

\def\href#1#2{#2}
\def\half{{1 \over 2}}

\newcommand{\ber}{\begin{eqnarray}}
\newcommand{\eer}{\end{eqnarray}}
\newcommand{\beqar}{\begin{eqnarray}}

\newcommand{\cR}{{\cal R}}

\newcommand{\eeqar}{\end{eqnarray}}

\newcommand{\dsl}
  {\kern.06em\hbox{\raise.15ex\hbox{$/$}\kern-.56em\hbox{$\partial$}}}

\newcommand{\eeqarr}{\end{eqnarray}}
\newcommand{\ZZ}{{\rm \kern 0.275em Z \kern -0.92em Z}\;}

\def\CC{{\mathchoice
{\rm C\mkern-8mu\vrule height1.45ex depth-.05ex
width.05em\mkern9mu\kern-.05em}
{\rm C\mkern-8mu\vrule height1.45ex depth-.05ex
width.05em\mkern9mu\kern-.05em}
{\rm C\mkern-8mu\vrule height1ex depth-.07ex
width.035em\mkern9mu\kern-.035em}
{\rm C\mkern-8mu\vrule height.65ex depth-.1ex
width.025em\mkern8mu\kern-.025em}}}

\def\RR{{\rm I\kern-1.6pt {\rm R}}}

\def\ZZ{{\rm Z}\kern-3.8pt {\rm Z} \kern2pt}
\def\IB{\relax{\rm I\kern-.18em B}}
\def\ID{\relax{\rm I\kern-.18em D}}
\def\II{\relax{\rm I\kern-.18em I}}
\def\IP{\relax{\rm I\kern-.18em P}}

\newcommand{\bear}{\begin{eqnarray}}
\newcommand{\eear}{\end{eqnarray}}

\newcommand{\x}{{\cal O}}

\def\to{\rightarrow}

\def\to{\rightarrow}

\def\A6{\mathcal{A}_6}
\def\ac{\mathfrak{a}^\c}
\def\am6{\mathfrak{a}^{\A6}}
\def\fz{\mathfrak{z}}

\def\c{\gamma}
\def\d{\delta}

\def\l{\lambda}
\def\m{\mu}
\def\n{\nu}
  
\def\r{\rho}
\def\vr{\varrho}   

\def\x{\xi}

\def\6{\partial}

\def\bea{\begin{eqnarray}}
\def\eea{\end{eqnarray}}

\def\beqx{\begin{displaymath}}
\def\eeqx{\end{displaymath}}

\newcommand{\bmat}{\left(\begin{array}}
\newcommand{\emat}{\end{array}\right)}

\def\half{\frac{1}{2}}

\def\c{\chi}
\def\d{\delta}

\def\l{\lambda}
\def\m{\mu}
\def\n{\nu}

\def\p{\pi}

\def\r{\rho}

\def\x{\xi}

\def\L{\Lambda}

\def\vr{\varrho}

\def\fz{\mathfrak{z}}

\def\bo{{\raise-.3ex\hbox{\large$\Box$}}}               
\def\face{{\raise.2ex\hbox{$\displaystyle \bigodot$}\mskip-2.2mu \llap {$\ddot
        \smile$}}}                                   
\def\>{\rangle}                                      
\def\<{\langle}                                      

\def\leftrightarrowfill{$\mathsurround=0pt \mathord\leftarrow \mkern-6mu
        \cleaders\hbox{$\mkern-2mu \mathord- \mkern-2mu$}\hfill
        \mkern-6mu \mathord\rightarrow$}        
\def\dvec#1{\vbox{\ialign{##\crcr
        \leftrightarrowfill\crcr\noalign{\kern-1pt\nointerlineskip}
        $\hfil\displaystyle{#1}\hfil$\crcr}}}           

\def\-{\hphantom{-}}
\newcommand{\dd}{\mbox{d}}

\begin{document}

\title{Bound states and deconfinement from Romans supergravity with magnetic flux}

\author{Ali Fatemiabhari\,\orcidlink{\orcidauthorFATEMIABHARI}}
\email{alifatemiabhari@gmail.com}
\affiliation{Department of Physics, Faculty  of Science and Engineering, Swansea University, Singleton Park, SA2 8PP, Swansea, Wales, United Kingdom}
\affiliation{Centre for Quantum Fields and Gravity, Faculty  of Science and Engineering, Swansea University, Singleton Park, SA2 8PP, Swansea, United Kingdom}

\author{Maurizio Piai\,\orcidlink{\orcidauthorPIAI}}
\email{m.piai@swansea.ac.uk}
\affiliation{Department of Physics, Faculty  of Science and Engineering, Swansea University, Singleton Park, SA2 8PP, Swansea, Wales, United Kingdom}
\affiliation{Centre for Quantum Fields and Gravity, Faculty  of Science and Engineering, Swansea University, Singleton Park, SA2 8PP, Swansea, United Kingdom}

\date{\today}

\begin{abstract}
We apply the dictionary of gauge-gravity dualities to study the spectrum of bound states in a special one-parameter family of strongly coupled, confining field theories in four dimensions. The top-down, holographic gravity dual description of this class of theories has been identified recently. It consists of non-supersymmetric regular background solutions of Romans half-maximal supergravity theory in six dimensions, in the presence of a non-trivial Abelian magnetic flux along a compactified direction of the geometry.  A zero-temperature, deconfinement, first-order phase transition appears at one end of this branch of solutions. It is  triggered by the strength of the flux, setting an upper bound on the magnitude of the magnetic flux that can be supported by the geometry.  We compute the spectrum of  fluctuations of the background fields in the gravity description, that correspond to field-theory bound states. Two scalar particles are the lightest in the spectrum, their masses being suppressed and almost degenerate across the whole parameter space.  Away from the transition, the heaviest between these two particles is identified as a dilaton, the pseudo-Nambu-Goldstone boson associated with scale invariance. It couples to the trace of the stress-energy tensor of the dual field theory, while the lightest scalar does not. In the range of parameter space closest to the extremum of the one-parameter family, near the first-order phase transition, a region with large curvature appears at the end of space of the geometry of the solutions. In this range, the two scalars mix non-trivially, and their masses are parametrically suppressed, in respect to the other bound states. 
\end{abstract}

\maketitle

\tableofcontents

\section{Introduction}
\label{Sec:introduction}

Gauge-gravity dualities (aka, holography)~\cite{Maldacena:1997re,Gubser:1998bc,Witten:1998qj,Aharony:1999ti} provide a powerful instrument for the study of non-perturbative phenomena in field theory. Special classes of strongly coupled confining gauge theories can be described by their gravity dual, characterised by the presence of an end of space in the holographic direction, in holographic correspondence to the confinement scale. The physics information related to confinement is encoded in the weakly curved, regular background geometry by the smooth shrinking of a portion of the higher dimensional space. For example, in Refs.~\cite{Witten:1998zw,Wen:2004qh,Kuperstein:2004yf,Brower:2000rp,Elander:2013jqa} the background geometries end when an internal circle decreses to zero size, and in Refs.~\cite{Chamseddine:1997nm,Klebanov:2000hb,Maldacena:2000yy,Butti:2004pk,Dymarsky:2005xt,Andrews:2006aw,Hoyos-Badajoz:2008znk,Nunez:2008wi,Elander:2009pk,Cassani:2010na,Bena:2010pr,Bennett:2011va,Dymarsky:2011ve,Maldacena:2009mw,Gaillard:2010qg,Caceres:2011zn,Elander:2011mh,Elander:2012yh,Elander:2017hyr,Elander:2017cle} the regular backgrounds contain a 2-cycle, related to the base of the conifold~\cite{Candelas:1989js,Klebanov:1998hh,Klebanov:2000nc,Papadopoulos:2000gj}, that collapses and disappears, yet without introducing singularities in the geometry. 

In the strong coupling regime of the field theory, physical observables can be computed in a prescriptive way, using the weakly curved gravity description and the well established dictionary relating the two sides of the correspondence. Holographic renormalisation is applied to compute the free energy, as well as  condensates and correlation functions involving field-theory local operators~\cite{Bianchi:2001kw,Skenderis:2002wp,Papadimitriou:2004ap}. This technique allows to establish the emergence of phase transitions, at finite as well as vanishing temperature and chemical potential (see the pedagogical discussion in Ref.~\cite{Casalderrey-Solana:2011dxg} and references therein, for example Refs.~\cite{Chamblin:1999tk,Chamblin:1999hg,Gubser:1998jb,Cai:1998ji,Cvetic:1999ne,Cvetic:1999rb,Kim:2006gp,Horigome:2006xu,Kobayashi:2006sb,Mateos:2007vc,Nakamura:2006xk, Karch:2007pd}). The characterisation of such transitions can be even used to study out-of-equilibrium phenomena, such as those governed by the dynamics of the bubbles arising in proximity of finite temperature phase transitions~\cite{Bigazzi:2020phm,Ares:2020lbt,Bea:2021zsu, Bigazzi:2021ucw,Henriksson:2021zei,Ares:2021ntv,Ares:2021nap,Morgante:2022zvc,Bea:2021zol, Bea:2022mfb,Bea:2024xgv,Bea:2024bxu,Bea:2024bls}, as the associated gravitational-wave emission~\cite{Witten:1984rs,Kamionkowski:1993fg,Allen:1996vm,Schwaller:2015tja,Croon:2018erz,Christensen:2018iqi}. Non-local operators, such as the Wilson loop, can also be studied, by including extended objects that probe the gravity background~\cite{Maldacena:1998im,Rey:1998ik}, hence demonstrating non-perturbative phenomena such as linear confinement---see for example Refs.~\cite{Brandhuber:1998bs,Brandhuber:1998er,Brandhuber:1999jr,Nunez:2009da}. The spectrum of bound states can be computed algorithmically as well,  for instance by using the gauge-invariant formalism developed in Refs.~\cite{Bianchi:2003ug,Berg:2005pd,Berg:2006xy,Elander:2009bm, Elander:2010wd,Elander:2010wn, Elander:2014ola,Elander:2018aub,Elander:2020csd}.  

Two complementary ways to exploit gauge-gravity dualities can be found in the literature. The  bottom-up approach led to useful phenomenological results and inspiration for new model-building ideas, by making use of simplified theories of gravity. One postulates the existence of a limited number of fields that propagate in extra dimensions, and assumes their dynamics to be  governed by a phenomenologically motivated classical action. The literature on this  approach to holography is vast and multifaceted, as is well exemplified by the extensive amount of work it stimulated on the special topic of the holographic dilaton~\cite{Goldberger:1999uk,DeWolfe:1999cp,Goldberger:1999un,Csaki:2000zn,Arkani-Hamed:2000ijo,Rattazzi:2000hs,Kofman:2004tk,Elander:2011aa,Kutasov:2012uq,Evans:2013vca,Hoyos:2013gma,Megias:2014iwa,Elander:2015asa,Megias:2015qqh,Athenodorou:2016ndx,Pomarol:2019aae,CruzRojas:2023jhw,Pomarol:2023xcc,CruzRojas:2025qcj}, the Pseudo-Nambu-Goldstone-Boson (PNGB) associated with the spontaneous breaking of approximate scale invariance~\cite{Coleman:1985rnk}. 

Conversely, by requiring a fundamental origin for the gravity theory, the top-down approach allows to gain access to more detailed information about the fundamental relation between different non-perturbative phenomena, because in this case also the field theory dual is better determined. The resulting increase in predictive power comes at the price of having to solve more complicated sets of coupled background equations. Avoiding the arising of singularities is challenging, as it requires imposing precise relations between different boundary conditions for the fields. In this more rigorous holographic framework, solution generating techniques, such as the one proposed in Refs.~\cite{Anabalon:2021tua} and exploited in Refs.~\cite{Nunez:2023xgl, Nunez:2023nnl,Fatemiabhari:2024aua,Chatzis:2024top,Chatzis:2024kdu,Kumar:2024pcz,Macpherson:2024qfi,Chatzis:2025dnu,Chatzis:2025hek,Anabalon:2025sok,Macpherson:2025pqi,Anabalon:2026yxk}, are an essential additional tool. They can be applied to the catalogue of known classical supergravity theories in various dimensions, exploiting their interconnection, through lifts and consistent truncations---see the lecture notes in Ref.~\cite{Samtleben:2008pe}, and references therein. Romans half-maximal supergravity theory in six dimensions plays a special role in this context~\cite{Romans:1985tw,DeWitt:1981wm,Giani:1984dw,Romans:1985tz}, thanks to the comparative simplicity of the boson contribution to its action, and the flexibility it offers to generalise the coset structure governing its sigma-model fields---see Refs.~\cite{
Brandhuber:1999np,Cvetic:1999un,Legramandi:2021aqv,Hong:2018amk,Jeong:2013jfc,DAuria:2000afl,Andrianopoli:2001rs,
Nishimura:2000wj,Gursoy:2002tx,Nunez:2001pt,Karndumri:2012vh,Lozano:2012au,Karndumri:2014lba,
Chang:2017mxc,Gutperle:2018axv,Suh:2018tul,Suh:2018szn,Kim:2019fsg,Chen:2019qib}.

Within the context of  the holographic description of confining gauge theories, this paper is part of a narrowly defined programme. We aim at exploiting the tools provided by gauge-gravity dualities in order to understand the relation between the presence (and nature) of phase transitions appearing at zero temperature, in the parameter space of a strongly coupled  field theory, and the spectrum of its bound states. One of the primary goals of this programme is to classify under what conditions a light dilaton appears in the spectrum. Furthermore, it is interesting to understand under what circumstances its mass can be (parametrically) dialled down, by moving in parameter space, relative to the position of the phase transitions. This programme can serve as foundational work for dilaton effective field theory (dEFT)~\cite{Goldberger:2007zk,Matsuzaki:2013eva,Golterman:2016lsd,Kasai:2016ifi,Hansen:2016fri,Golterman:2016cdd,Appelquist:2017wcg,Appelquist:2017vyy,Cata:2018wzl,Golterman:2018mfm,Cata:2019edh,Appelquist:2019lgk,Golterman:2020tdq,Golterman:2020utm, Appelquist:2022mjb,Zwicky:2023fay} and its applications~\cite{Appelquist:2020bqj,Appelquist:2022qgl,Cacciapaglia:2023kat,Appelquist:2024koa},  as it aims at identifying and characterising the conditions under which dEFT arises naturally from the underlying dynamics, giving a fundamental explanation for the size of its coefficients---see the discussion about dEFT power counting in Ref.~\cite{Appelquist:2022mjb}, for example. Scale invariance constraints the form of dEFT interactions~\cite{Coleman:1985rnk}, but determining the coefficients and scaling exponents requires additional input form the underlying theory. In the complementary context of fundamental field-theory studies, and in particular for  QCD-like theories, we refer the reader, for example, to Ref.~\cite{Migdal:1982jp} for Yang-Mills, as well as to the more recent Refs.~\cite{Zwicky:2023bzk,Zwicky:2023krx}, and the interesting comparison to lattice-QCD in Refs.~\cite{Stegeman:2025sca,Stegeman:2025tdl}.

Several examples of holographic theories relevant to these purposes have been presented in the literature. Broadly speaking, their physical properties can be classified in two main groups. In the presence of strong first-order phase transitions, as in the top-down constructions exhibited in Refs.~\cite{Elander:2020ial,Elander:2020fmv,Elander:2021wkc}, and the bottom-up models of Refs.~\cite{Elander:2022ebt,Fatemiabhari:2024lct}, a dilaton appears in the spectrum only along metastable branches of the gravity solution, that are not physically realised in the dual field theory. Conversely, if a line of first-order transitions ends at a critical point, in proximity of which the transition is weak (second order), then a parametrically light dilaton is part of the field-theory physical spectrum. This possibility has been demonstrated both in bottom-up~\cite{Faedo:2024zib}, and, more convincingly, in the top-down holographic context~\cite{Elander:2025fpk}.\footnote{For complementary approaches to a similar problem, see also the field theory study of special super-renormalisable lower-dimensional theories in Ref.~\cite{Cresswell-Hogg:2025kvr}, and the lattice field theories studied in Refs.~\cite{Lucini:2013wsa,Bennett:2022yfa}.} 

For this paper, we consider the one-parameter family of  backgrounds that have been presented in Ref.~\cite{Fatemiabhari:2024aua}. It is constructed by applying the procedure proposed in Ref.~\cite{Anabalon:2021tua} to Romans half-maximal supergravity theory in six dimensions~\cite{Romans:1985tw}. The background solutions admit a dual interpretation as confining field theories. A magnetic flux is present in the gravity background, providing a generalisation of the Melvin flux-tube solutions~\cite{Melvin:1963qx} studied in Refs.~\cite{Astorino:2012zm,Lim:2018vbq,Kastor:2020wsm}. We describe the construction in the body of the paper, by rewriting it after performing the dimensional reduction on a circle, down to five dimensions. We then report our results for the calculation of the holographically renormalised free energy and the field-theory spectrum of bound states.

We anticipate that the top-down holographic  theory discussed in this manuscript exhibits a more nuanced relation between phase transitions and the dilaton, with respect to the two classes mentioned above---see also Ref.~\cite{Piai:2026rst}. We find evidence of a first-order phase transition. We identify in the spectrum a particle with the properties of a dilaton, as well as a markedly suppressed mass. Yet,  we find that, atypically, the dilaton is not the lightest scalar state. Rather, it appears to be accompanied by a second, slightly lighter but almost mass degenerate, bound state, that has no relation to scale symmetry. In the region of parameter space near the first-order transition, both particles are significantly lighter than the rest of the bound states, and there is evidence of mixing between the two. The curvature of the background near the end of space grows when approaching the transition. These findings highlight how the nature of phase transitions in strongly coupled field theories, as well as the role that the spontaneous breaking of scale invariance plays in them, are open fields of investigation. Holographic techniques can lead to uncovering new, unexpected and non-trivial phenomena, worthy of further examination.

The paper is organised as follows. In Sect.~\ref{Sec:sugra} we describe the background gravity theory. We start by recalling, in Sect.~\ref{Sec:6}, the properties of Romans supergravity  in $D=6$ dimensions, which lifts to massive type IIA supergravity in $D=10$ dimensions. We perform its dimensional reduction on $S^1$, down to $D=5$ dimensions, and discuss it in Sect.~\ref{Sec:5}. We present the solutions of interest to this work in Sect.~\ref{Sec:solutions}. Our calculation of the free energy is exposed in Sect.~\ref{Sec:freeenergy}, and we discuss the scale setting procedure we adopt in Sect.~\ref{Sec:scale}. Our results for the spectra of fluctuations are presented  in Sect.~\ref{Sec:spectra}, separating the discussion of tensor, scalar, and vector modes in Sects.~\ref{Sec:spin2}, \ref{Sec:spin0}, and~\ref{Sec:spin1}, respectively. The spectrum of scalars is recalculated in the probe approximation, and the results are critically compared to those of the gauge-invariant calculation in Sect.~\ref{Sec:probe}. We conclude the paper with a summary and outlook section, that discusses future research avenues, in Sect.~\ref{Sec:outlook}. The general properties of the gauge-invariant formalism we adopt from the literature are reproduced in Appendix~\ref{Sec:sigma}. The numerical algorithm used to extract the spectra makes use of UV and IR expansions of the scalar fluctuations, that we report explicitly in Appendix~\ref{Sec:UVfluc}. 

\section{The supergravity theory}
\label{Sec:sugra}

In this section, we recall the main properties of the half-maximal supergravity theory in six dimensions that are relevant to this study, by restricting attention to the bosons. We perform a dimensional reduction of the theory to five dimensions, assuming the sixth dimension is compactified on a circle, $S^1$.  We then present and characterise the background solutions of interest.

\subsection{Romans half-maximal supergravity in $D=6$ dimensions} 
\label{Sec:6}

The half-maximal supergravity in $D=6$ dimensions presented by Romans in Ref.~\cite{Romans:1985tw}, can be obtained from the ten-dimensional massive type-IIA supergravity theory by warped reduction on $S^{4}$~\cite{Romans:1985tz}.  
We borrow notation from Ref.~\cite{Elander:2018aub} and references therein, adopt the mostly-plus signature
convention for the metric, and label six-dimensional quantities by hatted Roman indices, as $\hat{M} = 0,\, 1,\, 2,\, 3,\, 5,\, 6$. The bosonic part of the six-dimensional action contains 32 degrees of freedom. They can be listed as one scalar, $\phi$, the metric, $\hat g_{\hat{M}\hat{N}}$, a $U(1)$ vector, $A_{\hat{M}}$, with field strength $\hat F_{\hat{M}\hat{N}}$. Furthermore, there exists also  three vectors, $A_{\hat{M}}^{(i)}$, with $i=1,\, 2,\,3$,  transforming in the adjoint representation of $SU(2)$, associated to their field strengths, $\hat F_{\hat{M}\hat{N}}^{(i)}$. Finally, a 2-form, $B_{\hat{M}\hat{N}}$, is present, with its field strength, $\hat G_{\hat{M}\hat{N}\hat{T}}$. The topological terms involving $\hat F_{\hat{M}\hat{N}}$, $B_{\hat{M}\hat{N}}$, and $\hat F_{\hat{M}\hat{N}}^{(i)}$ are at least of the cubic order in the fields and are omitted in the following. They do not contribute to the relevant backgrounds and do not affect the (linearised) equations for the fluctuations. The action for the six dimensional theory can be written as in Ref.~\cite{Elander:2018aub}:\footnote{At variance with Ref.~\cite{Elander:2018aub}, we normalise the action with an additional multiplicative factor, $1/(2\pi)$, that does not affect any of the classical equations, but allows us to tighten the notation in later subsections. This scaling is equivalent to setting the Newton constant in six dimensions to $G_6 = \frac{1}{2}$, the action being written as $S_{6}=\frac{1}{16\pi\,G_6}\int \di^6x \sqrt{-\hat g_{6}}
\mathcal{R}_{6}+\cdots$.}
\begin{align}
S_{6}=\frac{1}{2\pi}\int_{}^{}\text{d}^{6}x\sqrt{-\hat g_{6}}
\bigg(\frac{\mathcal{R}_{6}}{4}&- \hat g^{\hat{M}\hat{N}}\partial_{\hat{M}}\phi\partial_{\hat{N}}\phi -\mathcal{V}_{6}(\phi)  
-\frac{1}{4}e^{-2\phi}\hat g^{\hat{M}\hat{R}}
\hat g^{\hat{N}\hat{S}}\sum_{i}\hat F_{\hat{M}\hat{N}}^{(i)}\hat F_{\hat{R}\hat{S}}^{(i)}\,+ \notag \\
&-\frac{1}{4}e^{-2\phi}\hat g^{\hat{M}\hat{R}}
\hat g^{\hat{N}\hat{S}}\hat{\mathcal{H}}_{\hat{M}\hat{N}}\hat{\mathcal{H}}_{\hat{R}\hat{S}}
-\frac{1}{12}e^{4\phi}\hat g^{\hat{M}\hat{R}}
\hat g^{\hat{N}\hat{S}}\hat g^{\hat{T}\hat{U}}\hat G_{\hat{M}\hat{N}\hat{T}}\hat G_{\hat{R}\hat{S}\hat{U}} \bigg)\,,
\label{ActionS6}
\end{align} 
with the explicit definitions
\begin{align}
\hat F_{\hat{M}\hat{N}}^{(i)} &\equiv \partial_{\hat{M}}A_{\hat{N}}^{(i)}-
\partial_{\hat{N}}A_{\hat{M}}^{(i)}+g\epsilon^{ijk}A_{\hat{M}}^{(j)}A_{\hat{N}}^{(k)}\,,\\
\hat F_{\hat{M}\hat{N}} &\equiv \partial_{\hat{M}}A_{\hat{N}}-
\partial_{\hat{N}}A_{\hat{M}}\,,\\
\hat{\mathcal{H}}_{\hat{M}\hat{N}} &\equiv \hat F_{\hat{M}\hat{N}} + mB_{\hat{M}\hat{N}}\,,\\
\hat G_{\hat{M}\hat{N}\hat{T}}&\equiv 3\partial_{\small[\hat{M}}B_{\hat{N}\hat{T}\small]}=
\partial_{\hat{M}}B_{\hat{N}\hat{T}}+
\partial_{\hat{N}}B_{\hat{T}\hat{M}}+
\partial_{\hat{T}}B_{\hat{M}\hat{N}}\,.
\end{align}
The anti-symmetrisation over indexes, denoted with $[\cdots]$, includes a normalisation,  with $[n_1n_2\cdots n_p]\equiv \frac{1}{p!}(n_1n_2\cdots n_p-n_2n_1\cdots n_p +\cdots)$. The  determinant of the metric is denoted as $\hat g_{6}$, while $\mathcal{R}_{6}$ is the corresponding Ricci scalar. The field-strength  {\small$\hat{\mathcal{H}}_{\hat{M}\hat{N}}$}  introduces a coupling between the $U(1)$ vector and 2-form fields, while {\small$\hat G_{\hat{M}\hat{N}\hat{T}}$} is the 2-form field strength tensor. Units are fixed in such  a way that the gauge coupling is $g=\sqrt{8}$. The mass parameter is chosen to be $m=\frac{2\sqrt{2}}{3}$. The scalar, $\phi$, has a potential, and with these conventions it reads as follows:
\be
\mathcal{V}_{6}(\phi)=\frac{1}{9}(e^{-6\phi}-9e^{2\phi}-12e^{-2\phi})\,.
\ee

\subsection{Reduction and truncation}
\label{Sec:5}

In order to make the exposition self-contained, we repeat here the results of Ref.~\cite{Elander:2020ial}, on the circle reduction along one of the external directions, leading to an effective description in five dimensions. The modulus controlling the circumference of the compact direction is promoted to a dynamical scalar field, $\chi$, in the reduced theory. The six-dimensional geometry is written using the metric ansatz
\begin{equation}
\text{d}s_{6}^{2}=e^{-2\chi} \text{d}s_{5}^{2} + e^{6\chi}
\big(\text{d}\eta + V_{M}\text{d}x^{M} \big)^{2}\,,    
\end{equation}
 where $\dd s_5^2$ is the generic five-dimensional metric, acting on the space with coordinates $X^M$, while $V_{M}$ is a five-dimensional vector field, the gravi-photon, that arises naturally from the reduction of the metric. Capital indices, $M=0,\, 1,\, 2,\, 3,\, 5$, label the non-compact directions, while $0\leq \eta <2\pi$ denotes the compact coordinate. In addition, the six-dimensional $SU(2)$ gauge fields are split according to  $A_{\hat{M}}^{(i)}=\{A_{\mu}^{(i)}, A_{5}^{(i)}, A^{(i)}_6\}$, with $\mu=0,\, 1,\, 2,\, 3$  restricted to the four-dimensional space-time indices.

We assume that the identities $\partial_6 \phi = \partial_{6}A_{\hat{N}} = \partial_{6}A^{(i)}_{\hat{N}} = \partial_{6}B_{\hat{N}\hat{T}} = 0$ hold exactly, which amounts to reducing the theory by retaining only the zero modes descending from the compactification on the circle. The action can be rewritten as 
\be
\label{eq:S6withtotalder}
S_{6}=\frac{1}{2\p}\int_{}^{}\text{d}\eta
\bigg\{S_{5}+\half \int_{}^{}\text{d}^{5}x\,
\partial_{M}\big(\sqrt{-g_{5}}\,g^{MN}\partial_{N}\chi\big)\bigg\},
\ee
where the total derivative term can be ignored in the following, as it does not enter the classical equations. The system is  governed by the five-dimensional gravity action, that  reads as follows:
{\small\begin{align}
S_{5}=\int_{}^{}\text{d}^{5}x\sqrt{-g_{5}}
\bigg(\frac{\mathcal{R}_{5}}{4}-\half G_{ab} g^{MN}\partial_{M}\Phi^{a}\partial_{N}\Phi^{b} &-\mathcal{V}(\phi,\chi)  
-\frac{1}{4}H_{AB}g^{MR}g^{NS}F_{MN}^{A}F_{RS}^{B}\,\notag\\
-\frac{1}{4}e^{2\chi-2\phi}g^{MR}g^{NS}\mathcal{H}_{MN}\mathcal{H}_{RS}
&-\frac{1}{12}e^{4\chi+4\phi}g^{MR}g^{NS}g^{TU}G_{MNT}G_{RSU}\,\notag\\
-\frac{1}{2}e^{-6\chi-2\phi}g^{NS}\mathcal{H}_{6N}\mathcal{H}_{6S}
&-\frac{1}{4}e^{-4\chi+4\phi}g^{NS}g^{TU}G_{6NT}G_{6SU}\,\notag\\
{-\frac{1}{2} e^{-6\chi-2\phi} g^{MN}(\partial_{M}A^{(i)}_6+g \epsilon^{ijk}A_{M}^{(j)}A_{6}^{(k)})}&{(\partial_{N}A^{(i)}_6+g \epsilon^{ijk}A_{N}^{(j)}A_{6}^{(k)})}\bigg)\,.\label{Eq:S5ast}
\end{align}}%
Notice that $A_6^{(i)}$ are three scalar fields, with $i=1,\,2,\,3$, and yet, at this stage, we treat them separately from the other sigma-model scalars. We will return to this point shortly. The potential is 
\beq
	\mathcal{V}(\phi,\chi) = e^{-2\chi}\mathcal{V}_{6}(\phi) \,.
\eeq
The sigma-model scalars, $\Phi^{a} = \{\phi,\, \chi\}$ and the field strengths, $\{F_{MN}^{V},\,F_{MN}^{(i)}\},$ have internal metric written as follows:
\beqs
	G_{ab} &=& \text{diag} \Big(2,\, 6 \Big) \,, \\
	 H_{AB} &=& \text{diag}\Big(\frac{1}{4}e^{8\chi},\,e^{2\chi-2\phi}\Big) \,,
\eeqs
where the field-strength tensors are defined as
\begin{align}
F_{MN}^{V}&\equiv \partial_{M}V_{N}-\partial_{N}V_{M}\,,\\
F_{MN}^{(i)}&\equiv \partial_{M}A_{N}^{(i)}-\partial_{N}A_{M}^{(i)}
+g\epsilon^{ijk}A_{M}^{(j)}A_{N}^{(k)}
+(V_{M}(\partial_{N}A^{(i)}_6+g \epsilon^{ijk}A_{6}^{(j)}A_{N}^{(k)})-V_{N}(\partial_{M}A^{(i)}_6+g \epsilon^{ijk}A_{M}^{(j)}A_{6}^{(k)}))\,.
\label{Eq:nA}
\end{align} 
The remaining tensor fields appearing in the five-dimensional action in Eq.~(\ref{Eq:S5ast}) are written as follows:
\begin{align}
\mathcal{H}_{MN}& \equiv \hat F_{MN}+mB_{MN} + \left( V_M \partial_N A_6 - V_N \partial_M A_6 \right) + m \left( B_{6M} V_N - B_{6N} V_M \right)\,,\\
\mathcal{H}_{6N}& \equiv \hat{\mathcal{H}}_{6N} = \partial_{6}A_{N}-\partial_{N}A_{6}+mB_{6N}=-\partial_{N}A_{6}+mB_{6N}\,,\\
G_{MNT}& \equiv 3\partial_{\small[M}B_{NT\small]}  - 6V_{\small[M} \partial_N B_{T\small]6}\,,\\
G_{6NT}&\equiv \hat G_{6NT} =
\partial_{6}B_{NT}-
\partial_{N}B_{6T}+
\partial_{T}B_{6N}=\partial_{T}B_{6N}-\partial_{N}B_{6T}\,.
\end{align}

The $32$ bosonic degrees of freedom present in the six dimensional theory are now decomposed as six scalar fields, the metric ($5$ d.o.f.), six vector fields ($3$ d.o.f. each), and one 2-form field ($3$ d.o.f.).

\subsubsection{Abelian truncation}
\label{Sec:Abelian}

We can truncate the six-dimensional 2-form, and the six-dimensional $U(1)$ vector, $A_{\hat{M}}$, that couples with it. In five-dimensional language, doing so amounts to removing the second and third line of Eq.~(\ref{Eq:S5ast}), and hence removing from the field content one scalar, two vector fields, and one 2-form.

We further truncate the theory by setting $F_{MN}^{(1,2)}=0=A_M^{(1,2)}=A_6^{(1,2)}$, hence restricting attention to the Abelian subgroup of $SU(2)$ associated with its third generator. The symmetries of the action ensure that this is a consistent truncation, yet we verified explicitly that this is the case; all the background solutions and classical fluctuations of the truncated system of equations are also solutions and fluctuations in the non-truncated one.  This further truncation leads to three main changes to the action, in Eq.~(\ref{Eq:S5ast}). Firstly,  the last line reduces to an additional entry that  extends the sigma-model kinetic term in the first line,  which now has fields $\Phi^{a} = \{\phi,\, \chi,\, A_6^{(3)}\}$, with  $3\times 3$ sigma-model metric
\beqs
	G_{ab} &=& \text{diag} \Big(2, \,6, \,{e^{-6\chi-2\phi}}\Big) \,.
\eeqs
Secondly,  in Eq.~(\ref{Eq:nA}) we can ignore the terms proportional to $\epsilon^{ijk}$, that encode the non-Abelian nature of $SU(2)$, by restricting the gauge index to $i=3$. Third and last, only $F_{MN}^{(3)}$ enters the last term of the first line of Eq.~(\ref{Eq:S5ast}), as we drop $F_{MN}^{(1,2)}=0$, though the metric in the $\{F_{MN}^{V},\,F_{MN}^{(3)}\},$ still looks the same:
\beqs
	 H_{AB} &=& \text{diag}\Big(\frac{1}{4}e^{8\chi},\,e^{2\chi-2\phi}\Big) \,.
\eeqs
We are hence left, in five dimensions,  with three scalars, $\Phi^{a} = \{\phi,\, \chi,\, A_6^{(3)}\}$, the metric tensor, and two vectors, $\{V_M,\, A^{(3)}_M\}$.  We have arrived to the considerably simpler action we use in the rest of the paper:
{\small\begin{align}
S_{5}^{\prime}=\int_{}^{}\text{d}^{5}x\sqrt{-g_{5}}
\bigg(\frac{\mathcal{R}_{5}}{4}-\half G_{ab} g^{MN}\partial_{M}\Phi^{a}\partial_{N}\Phi^{b} &-\mathcal{V}(\phi,\chi)  
-\frac{1}{4}H_{AB}g^{MR}g^{NS}F_{MN}^{A}F_{RS}^{B}\,\bigg)\,.\label{Eq:S5astprime}
\end{align}}%

\subsection{Background equations and solutions }
\label{Sec:solutions}

The background gravity solutions of interest take the domain-wall form in five dimensions, written as follows: 
\beqs
	\dd s_5^2 &=& \dd r^2 + e^{2A(r)} \dd x_{1,3}^2\,, 
 \eeqs
where the warp factor, $A$,  depends only on the holographic coordinate, $r$. The scalar fields,  $\phi$, $\chi$, and $A^{(3)}_6$,  retained in Sect.~\ref{Sec:Abelian} are also non-trivial, having a $r$-dependent background profile, while the vectors, $V_M$ and $A^{(3)}_M$, vanish. The  metric in five dimensions can be rewritten in terms of a new $\rho$ coordinate as
\beqs
\di s_5^2&=&e^{2A(\rho)}\di s_{1,3}^2 +e^{2\chi(\rho)} \di \rho^2\,.
\eeqs
The lift of the  background metric to six dimensions takes the form
\beqs
	\dd s_6^2=  \dd \rho^2 + e^{2A(\rho)-2\chi(\rho)} \dd x_{1,3}^2 + e^{6\chi(\rho)} \dd \eta^2\,.
\eeqs

The equations of motion, obtained with this ansatz from the five dimensional action, and including also Einstein equations, are given by the following system of  coupled, non-linear, second-order, ordinary differential equations, supplemented by a constraint:
\beqs
\partial_{\rho}^2\phi +(4\partial_{\rho}A- \partial_{\rho}\chi)\partial_{\rho}\phi+\frac{1}{2} e^{-6 \chi-2\phi}\left(\partial_{\rho}A^{(3)}_6\right)^2&=\frac{1}{2}\frac{\partial \mathcal{V}_6}{\partial \phi}\,,
\label{Eq:Eqrho0}\\
\partial_{\rho}^2\chi +(4\partial_{\rho}A-\partial_{\rho}\chi)\partial_{\rho}\chi+\frac{1}{2} e^{-6 \chi-2\phi}\left(\partial_{\rho}A^{(3)}_6\right)^2&=-\frac{1}{3}\mathcal{V}_6\,,
\label{Eq:Eqrho1}
\\
\label{Eq:Eqrho2}
\partial_{\rho}^2 A^{(3)}_6+(4\partial_{\rho}A-7\partial_{\rho}\chi-2\partial_{\rho} \phi )\partial_{\rho}A^{(3)}_6&=0\,,\\
3\partial_{\rho}^2A +6(\partial_{\rho}A)^2+2(\partial_{\rho} \phi)^2+6(\partial_{\rho}\chi)^2 + e^{-6\chi-2\phi}\left(\partial_{\rho}A^{(3)}_6\right)^2-3\partial_{\rho}\chi\partial_{\rho}A&=-2\mathcal{V}_6\,,
\label{Eq:Eqrho3}
\\
\label{Eq:Eqrho4}
6(\partial_{\rho}A)^2-6(\partial_{\rho}\chi)^2 -2(\partial_{\rho} \phi)^2-e^{-6\chi-2\phi}\left(\partial_{\rho}A^{(3)}_6\right)^2&=-2\mathcal{V}_6\,.
\eeqs

Before providing the solutions of interest, we notice that close inspection of this differential system leads to identifying a conserved quantity. From the observation that Eq.~(\ref{Eq:Eqrho2}) can be written as a total derivative, one deduces that the associated conserved quantity is
\begin{equation}\label{eq:conserved quantity}
   C_{A} \equiv e^{4A-7\chi-2\phi}\partial_{\r}A^{(3)}_6\,,
\end{equation} 
that obeys $\partial_{\rho} C_{A} =0$, when evaluated on the backgrounds.

All solutions relevant for our analysis share a common asymptotic form in the large-$\rho$ region of the geometry, in which the six-dimensional geometry is approaching AdS$_6$. In field theory terms, this corresponds to the expectation that the ultraviolet (UV) regime in the dual field theory is described by a deformation of a strongly coupled conformal field theory in five dimensions. It is convenient to introduce a third alternative parameterisation of the holographic coordinate, $z \equiv e^{-2/3\rho}$, so that the asymptotic regime is given by the $z\rightarrow 0$ limit. In terms of this variable, the background configurations can be expressed in terms of a UV expansion, organised as a power series in the small quantity, $z \ll 1$.
The UV expansion of the background functions takes the following form:
\beqs
\phi(z)&=&\phi_\alpha z^2 +\phi_\beta z^3 -6 \phi_\alpha ^2 z^4  -4 \phi_\alpha  \phi_\beta  z^5 +\frac{1}{2} z^6 \left(29 \phi_\alpha^3-2 \phi_\beta^2\right)+\,{\cal O}(z^{7})\,,\\
A^{(3)}_6 (z)&=&A^{(3),U}_6\,+\,A^{(3),3}_6 z^3\,+\,
\frac{6}{5}\phi_\alpha\,A^{(3),3}_6 z^5\,+\, \phi_\beta A^{(3),3}_6 z^6\,+\,{\cal O}(z^{7})\,,\\
\chi(z)&=&\chi^{U}
-\frac{1}{3}\log(z)\,-\,\frac{1}{12} \phi_\alpha ^2 z^4 \,+\,\chi^{(5)} z^5\,-\,\frac{1}{36}\left(32 \phi_\alpha ^3-3 \phi_\beta ^2\right) z^6 +\,{\cal O}(z^{7})\,,\\
A(z)&=&A^{U}-\frac{4}{3}\log(z)\,-\,\frac{1}{3} \phi_\alpha ^2 z^4 \,+\,\frac{1}{20} z^5 \left(5 \chi ^{(5)}-12 \phi_\alpha  \phi_\beta \right)\,+\,\frac{1}{9} z^6 \left(32 \phi_\alpha ^3-3 \phi_\beta ^2\right)
\,+\,{\cal O}(z^{7})\,.
\eeqs
Here, $A^{U}$, and $\chi^{U}$ play the role of trivial integration constants. The additional parameters, $A^{(3),U}_6$, $\chi^{(5)}$, $\phi_\alpha$, $\phi_\beta$, and $A^{(3),3}_6$, correspond to field-theory quantities.

For this study, we focus on two classes of exact solutions, both known in closed form, the UV expansions of which are distinguished by the form of $\chi^{(5)}$ and $A_6^{(3)}$. The first such class is obtained by setting $\partial_{\rho} A_6^{(3)}=0$, as well as  $A_6^{(3),3} = \phi_\alpha = \phi_\beta = \chi^{(5)} = 0 $. Thus, the first set of solutions of interest reads
\beqs
\phi(\rho)&=&0\,,\\
A^{(3)}_6 (\rho)&=&A^{(3),U}_6\,,\\
A(\rho)&=&A^{U}+\frac{8}{9}\rho\,,\\
\chi(\rho)&=&\chi^{U}+\frac{2}{9}\rho\,.
\eeqs
In this solution, the six dimensional metric is compatible with the  domain-wall ansatz, and takes the form
\beqs
	\dd s_6^2=  \dd \rho^2 + e^{2 \rho} \dd x_{1,3}^2 + e^{2\rho} \dd \eta^2\,,
\eeqs
which has AdS$_6$ geometry with unit curvature, although one dimension has been compactified on a circle. 
We devote the next subsection to a second, more interesting one-parameter family of solutions.

\subsubsection{Solutions with Abelian flux}

\begin{figure}[t]
\centering
\includegraphics[width=0.6\textwidth]{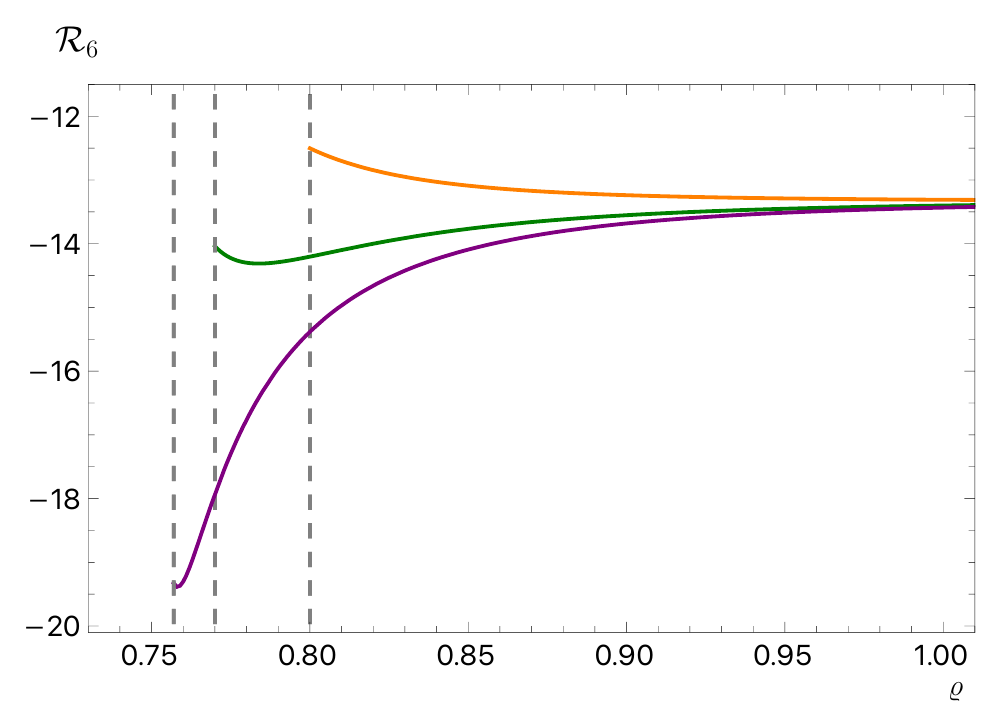}        
   \hfill
	\caption{The Ricci scalar, $\mathcal{R}_{6}$,  as a function of $\vr>\vr_0$, for representative examples of background solutions with non-trivial flux, obtained with the choices $\vr_0=0.757,\, 0.77,\, 0.8$, shown  in purple, green and orange, respectively. The vertical, dashed lines represent the end of space of the geometry, at $\vr=\vr_0$.
  \label{fig:Ricci}}
\end{figure}

The second class of solutions, introduced in Ref.~\cite{Fatemiabhari:2024aua}, has close similarities with the backgrounds described in Ref.~\cite{Anabalon:2021tua}. There are nontrivial values for the parameters $\chi^{(5)}$ and $A_6^{(3),3}$, appearing in the UV expansions.  In order to write these solutions in closed form, it is convenient to  introduce a fourth parametrisation of the  radial coordinate.  The metric in six dimensions, written in this new coordinate, takes the following form~\cite{Fatemiabhari:2024aua}:
\begin{equation}\label{eq: AR metric}
ds_6^{2}=H^{1/2}(\vr)\vr^{2}(d x_{1,3}^2)+\frac{H^{1/2}(\vr)d\vr^{2}}{f(\varrho)}+\frac{f(\vr)}{H^{3/2}(\vr)}d\eta^{2}.
\end{equation}
Comparison with the expressions derived from the five-dimensional language leads to the identifications
\beqs
H^{1/4}(\vr)\varrho&\equiv & e^{A(\rho)-\chi(\rho)}\,,\\
\frac{f(\varrho)}{H^{3/2}(\vr)}&\equiv & e^{6\chi(\rho)}\,,\\
\frac{\dd\varrho}{\dd \rho}&\equiv & \frac{\sqrt{f(\varrho)}}{H^{1/4}(\vr)}.
\eeqs
Equivalently, these expressions can be recast as $\chi(\rho)=\frac{1}{6}\log(\frac{f(\varrho)}{H^{3/2}(\vr)})$,
$A(\rho)=\frac{1}{6}\log(f(\varrho))+\log\left(\varrho\right)$,
and $\frac{\partial}{\partial \rho}= \frac{\sqrt{f(\varrho)}}{H^{1/4}(\vr)}\frac{\partial}{\partial \varrho}=\frac{1}{\vr^2}e^{2A(\vr)+\chi(\vr)}\frac{\partial}{\partial \vr}$.
The classical background equations, rewritten in this parametrisation of the holographic direction, are the following:
\beqs
\frac{1}{2} \frac{\partial \mathcal{V}_6}{\partial \phi}
&=&
\frac{1}{2} e^{-2 \phi (\vr)} H(\vr) \left(A^{(3) \prime}_6(\vr)\right)^2+\frac{4 f(\vr) \phi '(\vr)}{\vr\sqrt{H(\vr)}}+\frac{f'(\vr) \phi '(\vr)}{\sqrt{H(\vr)}}+\frac{f(\vr) \phi ''(\vr)}{\sqrt{H(\vr)}}\,, \\
-4 \mathcal{V}_6
&=&
6 e^{-2 \phi (\vr)} H(\vr) \left(A^{(3) \prime}_6(\vr)\right)^2+\frac{3 f(\vr) \left(H'(\vr)\right)^2}{H(\vr)^{5/2}}+\frac{\frac{8 f'(\vr)}{\vr}+2 f''(\vr)}{\sqrt{H(\vr)}}\\
&&
\nonumber
-\frac{3 \left(\vr f'(\vr)H'(\vr)+f(\vr) \left(4 H'(\vr)+\vr H''(\vr)\right)\right)}{\vr H(\vr)^{3/2}}\,,\\
0
&=&
3 \vr A^{(3) \prime}_6(\vr) H'(\vr)+H(\vr)\left(A^{(3) \prime}_6(\vr) \left(8-4 \vr \phi '(\vr)\right)+2 \vr A^{(3) \prime\prime}_6(\vr)\right)\,,\\
-2 \mathcal{V}_6
&=&
e^{-2 \phi (\vr)} H(\vr) \left(A^{(3) \prime}_6(\vr)\right)^2-\frac{f'(\vr) H'(\vr)}{2 H(\vr)^{3/2}}
+\frac{{6\, f'(\vr)}{}+{\vr\,f''(\vr)}{}}{2\vr\,\sqrt{H(\vr)}}\\
&&
\nonumber
+\frac{f(\vr) \left(3 \left(H'(\vr)\right)^2+H(\vr)^2 \left(\frac{24}{\vr^2}+16 \left(\phi '(\vr)\right)^2\right)\right)}{8 H(\vr)^{5/2}}
\,, \\
-2 \mathcal{V}_6&=&
-e^{-2 \phi (\vr)} H(\vr) \left(A^{(3) \prime}_6(\vr)\right)^2+\frac{2
   f'(\vr)}{\vr \sqrt{H(\vr)}}+\frac{f'(\vr) H'(\vr)}{2 H(\vr)^{3/2}}-\frac{3 f(\vr)\left(H'(\vr)\right)^2}{8 H(\vr)^{5/2}}
   \\
   &&
   \nonumber
   -\frac{2 f(\vr)\left(-3+\vr^2 \left(\phi '(\vr)\right)^2\right)}{\vr^2 \sqrt{H(\vr)}}\,,
\eeqs
where the primes denote derivatives taken with respect to $\varrho$.

The closed form of these soliton solutions takes the appealing, simple form:
\beqs
H(\vr)&=&1-\frac{c^2}{\vr^3}\,,\\
f(\vr)&=&\frac{4}{9} \vr^2 H(\vr)^2-\frac{\mu }{\vr^3}\,,\\
\phi(\vr)&=&-\frac{1}{4}\log(H(\vr))\,,\\
A^{(3)}_6&=&\frac{\sqrt{\mu}}{\sqrt{2}c}\left(  \frac{1}{H(\vr_0)}-\frac{1}{H(\vr)}\right)\,.
\eeqs
Here, $\vr_0$ is the largest positive root of $f(\vr_{0})=0$, with $\mu$ and $c$ two independent integration constants. Notice that an additive integration constant appearing in $A^{(3)}_6(\vr)$ has been set so that $A^{(3)}_6(\vr_0)=0$. This condition is necessary as the space ends at $\vr_0$, at which point the circle in the geometry is shrinking to zero size, and hence it cannot support a non-trivial flux. 
Along a similar line of reasoning, there is also the potential for a conical singularity appearing at the end of  the geometry, which we need to avoid. Hence, by expanding close to the end of space, at $\vr\simeq vr_0$, in the $\vr-\eta$ subspace, direct calculation of the induced metric demonstrates that one must impose the constraint
\beqs
\frac{f'(\vr_0)}{H(\vr_0)}&=& 2\,,
\eeqs
in order to avoid the appearance of such a singular behaviour. Doing so introduces a (non-trivial) relation between the parameters $c$ and $\mu$, demonstrating that the solutions of interest constitute a one-parameter class of backgrounds. We find it convenient to use $\vr_0$ to parametrise the resulting family of backgrounds, and write
\beqs \label{eq:cmucons}
\m=\vr_0^3 (-3 + 4 \vr_0)^2\,,\\
c=\pm \sqrt{9 \vr_0^2 - 10 \vr_0^3}/\sqrt{2}\,.
\eeqs
The ambiguity in the sign of $c$ reflects the fact that given any solution of the classical equation, another solution can be written by just changing the sign of $A^{(3)}_6$. As can be seen in Fig.~\ref{fig:Ricci}, in which we show the Ricci scalar associated with  the six-dimensional metric, all these solutions are regular, provided $\vr_0>3/4$. Similar results hold for other (higher-order) invariants, computed in six dimensions. The limiting case $\vr_0=\frac{3}{4}$ has a singularity located at the end of the space.

We find it useful to characterise  also the IR asymptotic behaviour of the backgrounds,  by writing a power expansion performed in proximity of $\vr\sim \vr_0$. What results is the following:
\begin{align}
    \phi(\vr)&=\frac{1}{4} \log \left(\frac{2 \vr_0}{12 \vr_0-9}\right)+\frac{(9-10 \vr_0) (\vr-\vr_0)}{12 \vr_0-16 \vr_0^2}-\frac{3 ((2 \vr_0-1) (10 \vr_0-9)) (\vr-\vr_0)^2}{8 \left((3-4 \vr_0)^2 \vr_0^2\right)}+{\cal O}((\vr-\vr_0)^3)\,,\\
    A_6^{(3)}(\vr)&=\frac{2 (2 \vr_0-3) \sqrt{\frac{9}{\vr_0}-10} (\vr-\vr_0)^2}{3 (3-4 \vr_0)^2}+\frac{2 \sqrt{(9-10 \vr_0) \vr_0} (\vr-\vr_0)}{9-12 \vr_0}+{\cal O}((\vr-\vr_0)^3)\,,\\
    \chi(\vr)&=\frac{1}{12} \left(2 \log (\vr-\vr_0)+\log \left(\frac{8
   \vr_0}{-9+12 \vr_0}\right)\right)\nonumber\\
   &+\frac{\left(-9-30 \vr_0+40 \vr_0^2\right) (\vr-\vr_0)}{36 \vr_0 (-3+4 \vr_0)}+\frac{\left(-81+684 \vr_0-1428 \vr_0^2+1600 \vr_0^3-800 \vr_0^4\right) (\vr-\vr_0)^2}{216 \vr_0^2 (-3+4 \vr_0)^2}+{\cal O}((\vr-\vr_0)^3)\,,\\
    A(\vr)&=\frac{1}{6} (\log (3\vr_0^5(-3+4\vr_0))+\log (\vr-\vr_0))+\frac{\left(-9+6 \vr_0+10 \vr_0^2\right) (\vr-\vr_0)}{9 \vr_0 (-3+4 \vr_0)}\nonumber\\
   &+\frac{\left(-81+252 \vr_0-384 \vr_0^2+400 \vr_0^3-200 \vr_0^4\right) (\vr-\vr_0)^2}{54
   \vr_0^2 (-3+4 \vr_0)^2}+{\cal O}((\vr-\vr_0)^3)\,.
\end{align}
In these expressions, we notice that the singular behaviour appearing in the limit $\vr_0\rightarrow 3/4$ affects also the scalar, $\phi$.

It is also useful to compare the asymptotic UV behaviour of the backgrounds in this class of solutions to the general  power series expansion in the small  $z=e^{-2/3\rho}$ defined earlier in the paper. The expressions for all the non-trivial background functions take the following form: 
\beqs 
\label{eq:uvexpansion1}
\phi(z)&=&\frac{c^2}{4}z^3-\frac{c^4 z^6}{16}-\frac{27 c^2 \mu }{160} z^8+\frac{19 c^6 }{768}z^9 +{\cal O}(z^{10})\,,\\
\label{eq:uvexpansion2}
A_6^{(3)}(z)&=& -\frac{c \sqrt{\mu }}{\sqrt{2} \left(c^2-\vr_0^3\right)}-\frac{c \sqrt{\mu}}{\sqrt{2}}z^3-\frac{\left(c^3 \sqrt{\mu }\right) z^6}{4 \sqrt{2}}+{\cal O}(z^{8})\,,\\
\label{eq:uvexpansion3}
   \chi(z)&=&\frac{1}{3}\log \left(\frac{2}{3}\right)
  - \frac{1}{3}\log \left(z\right)
   -\frac{3 \mu}{10} z^5-\frac{c^4}{192}  z^6+{\cal O}(z^{7})\,,\\
   \label{eq:uvexpansion4}
   A(z)&=&\frac{1}{3}\log \left(\frac{2}{3}\right)-\frac{4}{3} \log \left({z}\right) -\frac{3 \mu}{40} z^5-\frac{c^4}{48}  z^6+{\cal O}(z^{7})\,.
\eeqs
By comparing with the most general UV expansion, we arrive at the following identifications:
\beqs
\phi_\alpha&=&0\,, \\
 \phi_\beta &= &\frac{c^2}{4}\,,\\
A_6^{(3),3}&=&-\frac{c \sqrt{\mu}}{\sqrt{2}}\,,\\
A_6^{(3),U}&=&-\frac{c \sqrt{\mu }}{\sqrt{2} \left(c^2-\vr_0^3\right)}\,,\\
\chi^{(5)}&=&-\frac{3 \mu}{10}\,,\\
\chi^U&=&\frac{1}{3}\log \left(\frac{2}{3}\right)\,, \\
A^{U} &=& \frac{1}{3}\log \left(\frac{2}{3}\right)\,.
\eeqs

One can use this expansion for the fields near the boundary, at $z\to 0$, to study the deformations in the UV regime of the dual field theory. The QFT data needed are the operators turned on in the field theory dual, and their vacuum expectation value. In the expansions provided in Eqs.(\ref{eq:uvexpansion1})-(\ref{eq:uvexpansion4}), the non-normalizable (leading) modes are identified as the sources that couple to gauge-invariant operators in the boundary QFT, while the conjugate normalisable (subleading) modes encode the one-point functions, or vacuum expectation values (VEVs), of those operators. The normalisable modes—after removing divergences through local covariant counterterms—provide the renormalised VEVs in the presence of the sources. For example, turning on a constant Abelian gauge field along a compact, (Euclidean) time-like direction, in the gravity description, would have a natural interpretation in field theory as the presence of a non-zero chemical potential, the Lagrange multiplier that imposes the presence of a finite charge density in the dual field theory. 

In the case at hand, the presence of a constant, non-normalisable, leading term in the expansion of $A_6^{(3)}$, can be interpreted in the field theory as a Wilson line along the compact, space-like dimension. This being the double-Wick rotation of a chemical potential, its effect is to impose the presence of a current density in the field theory. In fact, for the solution studied here, one can deduce from the subleading terms of the expansions of $A_6^{(3)}$ and $\phi$ that two non-trivial VEVs, both for objects of dimension $\Delta=3$, are present: 
\begin{equation} \label{boundaryXA}
    \langle J \rangle = -\frac{c\sqrt{\mu} }{\sqrt{2}} , \quad  \langle {\cal O}_\phi \rangle = \frac{c^2}{4}\,.
\end{equation}
A more  detailed analysis is reported in Appendix B of Ref.~\cite{Fatemiabhari:2024aua}. As anticipated, $A_6^{(3)}$ acts as a source that enforces the presence of a non trivial current associated to the  global (Abelian) R-symmetry current $J$ at the boundary---see also the discussions in Refs.~\cite{Kumar:2024pcz,Castellani:2024ial}, and the precursor in Ref.~\cite{Cassani:2021fyv}. As we shall see later in the paper, the global stability analysis shows the presence of a zero-temperature quantum transition that is triggered by dialing the source in the field theory to large enough values, leading to deconfinement. Hence, there is a maximum allowed value of the flux, below which the field theory confines, and above which the field theory is deconfined.

\section{Global stability analysis: the free energy}
\label{Sec:freeenergy}

This section is devoted to the calculation of the free energy of the field theory associated with the classical background solutions. This is obtained by evaluating on shell the six-dimensional action in Eq.~(\ref{ActionS6}). To this purpose, we regulate the calculation by introducing two end points in the holographic direction, one in the region corresponding to the IR of the field theory, $\vr_1>\vr_0$, and one in the asymptotic region, corresponding to the UV, $\vr_2\gg \vr_1$. We then compute the action by restricting the integrals to the range $\vr_1<\vr<\vr_2$. These unphysical regulators must  be removed from the final results, by taking the limits $\vr_1 \to \vr_0$ and $\vr_2 \to \infty$, respectively. The introduction of  hyper-surfaces in correspondence to the regulators necessitates of the inclusion of a Gibbons–Hawking–York (GHY) boundary term, as well as boundary-localised potential terms, denoted by $\lambda_i$. They render well-defined  the variational problem yielding the classical equations and boundary conditions~\cite{Elander:2010wd}. They also act as counterterms in the field theory, by cancelling possible divergencies and implementing holographic renormalisation~\cite{Bianchi:2001kw,
 Skenderis:2002wp,Papadimitriou:2004ap}. 
 
 The on-shell action entering the free-energy calculation is therefore given by the following expression, which depends only on the background values of the metric and scalar fields $\chi$, $\phi$, and $A_6^{(3)}$ (see also Ref.~\cite{Elander:2020ial}):
\beqs
    \mathcal{S}&=&\mathcal{S}_6+\sum_{i=1, 2}(\mathcal{S}_{\mathcal{K},i}+\mathcal{S}_{\mathcal{\l},i})\\ \nonumber
    &=& \frac{1}{2\pi}\int_{\vr_1}^{\vr_2} \dd \eta\dd^4xd\vr  \sqrt{-\hat g_6}  \bigg\{ \frac{\mathcal R_6}{4} 
    - \hat g^{\hat{M}\hat{N}}\partial_{\hat{M}}\phi\partial_{\hat{N}}\phi -\mathcal{V}_{6}(\phi)  
-\frac{1}{2}e^{-2\phi}\hat g^{66}\hat g^{\hat{M}\hat{N}}
\partial_{\hat{M}} A_6^{(3)}\partial_{\hat{N}} A_6^{(3)}\bigg\} 
	\\
	\nonumber
	&&+ \sum_{i=1, 2}\frac{(-1)^i}{2\pi}\int \dd^4x d\eta \sqrt{-\tilde{g}}\left.\left(\frac{\mathcal{K}}{2}+\l_i\right)\right|_{\vr=\vr_i}\,,
\eeqs
In this expression, $\tilde{g}=-\vr_i^8f(\vr_i)\sqrt{H(\vr_i)}$ is the determinant of the induced metric evaluated at the boundary, $i=1,2$. The extrinsic curvature scalar, $\mathcal{K}$, is defined as:
\begin{equation}
        \mathcal{K}\equiv\hat g^{\hat M \hat N}\mathcal{K}_{\hat M \hat N}\equiv g^{\hat M \hat N}\nabla _{\hat M}n_{\hat N}.
\end{equation}
Here, $n_{\hat N}=(0, 0, 0, 0, \frac{H^{1/4}(\vr)}{\sqrt{f(\vr)}}, 0)$ denotes the normal boundary vector.  The covariant derivative  acts as $\nabla_{\hat  M}f_{\hat N}=\6_{\hat M}f_{\hat N} -\Gamma^{\hat P}_{\,\,\hat M \hat N}f_{\hat P}$, and the connection takes the form (see also Appendix~\ref{Sec:gaugeinvariantformalism}):
\be
\Gamma^{\hat P}_{\,\,\hat M \hat N}\equiv\frac{1}{2}g^{\hat P \hat Q}\left(\6_M \hat g_{\hat N \hat Q}+\6_{\hat N}\hat g_{\hat Q \hat M}-\6_{\hat Q}\hat g_{\hat M \hat N}\right).
\ee
With these conventions, the extrinsic curvature reads:
\begin{equation}
    \mathcal{K}=\frac{2 \vr H(\vr) f'(\vr)+f(\vr) \left(\vr H'(\vr)+16 H(\vr)\right)}{4 \vr \sqrt{f(\vr)} H(\vr)^{5/4}}\,.
\end{equation}

After imposing the equations of motion, and integrating over the angle, $\int \dd \eta =2\pi$, the action in the bulk  can be rewritten as a total derivative:
\begin{equation}
    \mathcal{S}_6=\int_{\vr_1}^{\vr_2}  \dd^4x\dd\vr \left[-\partial_{\vr}\left(\frac{\vr^3 f(\vr) \left(\vr H'(\vr)+4 H(\vr)\right)}{8 H(\vr)}\right)\right]\,,
\end{equation}
so the boundary contribution to the action is given by
\begin{equation}
    \mathcal{S}_i=(-1)^i\int \dd^4x \left.\left[\vr^4\sqrt{f(\vr)}H(\vr)^{1/4}\left(\frac{2 \vr H(\vr) f'(\vr)+f(\vr) \left(\vr H'(\vr)+16 H(\vr)\right)}{8 \vr \sqrt{f(\vr)} H(\vr)^{5/4}}+\l_i\right)\right]\right|_{\vr=\vr_i}\,.
\end{equation}
The complete on-shell action, ${\cal S}$, reads as follows:
\beqs
    \mathcal{S}&=&\int \dd^4x \left\{ \left.\left[-\frac{\vr^3 f(\vr) \left(\vr H'(\vr)+4 H(\vr)\right)}{8 H(\vr)}+\vr^4\sqrt{f(\vr)}H(\vr)^{\frac{1}{4}}\left(\frac{2 \vr H(\vr) f'(\vr)+f(\vr) \left(\vr H'(\vr)+16 H(\vr)\right)}{8 \vr \sqrt{f(\vr)} H(\vr)^{\frac{5}{4}}}+\l_2\right)\right]\right|_{\vr=\vr_2}\right.\nonumber\\
  &-& \left. \left.\left[-\frac{\vr^3 f(\vr) \left(\vr H'(\vr)+4 H(\vr)\right)}{8 H(\vr)}+\vr^4\sqrt{f(\vr)}H(\vr)^{\frac{1}{4}}\left(\frac{2 \vr H(\vr) f'(\vr)+f(\vr) \left(\vr H'(\vr)+16 H(\vr)\right)}{8 \vr \sqrt{f(\vr)} H(\vr)^{\frac{5}{4}}}+\l_1\right)\right]\right|_{\vr=\vr_1}\right\}\,.
\eeqs

One can define the free energy, $F$, and free energy density, $\mathcal{F}$, as
\beqs
    F=-\lim_{\vr_1 \rightarrow \vr_0} \lim _{\vr_2 \rightarrow \infty} \mathcal{S} \equiv \int\dd^4 x \, \mathcal{F}\,.
\eeqs
In order to be able to take the limits, we must specify the localised potentials. We choose $\lambda_1=-\frac{3}{2}\partial_{\rho}A(\r)$. This term is fixed by the requirement of having a well-posed variational principle. In particular, varying the bulk action together with the infrared boundary contribution produces the bulk equations of motion, along with the corresponding boundary conditions to be imposed at $\rho=\rho_1$. With this choice the boundary-localised contribution combines with the bulk contribution,  so that there is no spurious dependence of the free energy in the regulator at $\rho_1$. 

For the second boundary, holographic renormalisation~\cite{Bianchi:2001kw,
 Skenderis:2002wp,Papadimitriou:2004ap}  prescribes to set $\lambda_2={\cal W}_2$, where ${\cal W}_2$ is the superpotential of the theory in six dimensions, which can be written as a power expansion in $\phi$---see in Eq.~(34) of Ref.~\cite{Elander:2020ial}---to take the form
\beqs
{\cal W}_2&=&
-\frac{4}{3}\,-\,\frac{4}{3}\phi^2\,+\,{\cal O}(\phi^3)\,.
\eeqs
Higher-order  terms do not contribute to the final result for ${\cal F}$, in the limit $\vr_2\rightarrow +\infty$. 
It is worth mentioning that in general the cancellation of divergences performed introduces an inherent scheme dependence (see also Ref.~\cite{Bobev:2013cja}). Although it yields a finite expression, it also implies that the resulting free energy does not satisfy the assumptions required for the standard thermodynamic concavity theorems to hold.

After substituting the expressions for $f(\vr)$ and $H(\vr)$ and performing the limits one arrives to our final expression:
\beqs
    \mathcal{F}&=&-\lim _{\vr_2\rightarrow\infty}\left[\left.\frac{1}{12} \vr^3 \left(3 \vr f'(\vr)+\vr \sqrt{f(\vr)} \sqrt[4]{H(\vr)} \left(\log ^2(H(\vr))-16\right)+18 f(\vr)\right)\right|_{\vr=\vr_2}\right]\\
    &=& -\frac{\m}{4}\,=\,-\frac{1}{4}\vr_0^3 (-3 + 4 \vr_0)^2\,.
\eeqs
We note that, despite the presence of flux for the gauge fields, this expression is also in agreement with Eq.~(118) of Ref.~\cite{Elander:2020ial}, which reads:
 \beqs
 \label{Eq:ff}
 {\cal F} &=&-\left.\lim_{\rho_2\rightarrow \infty}
 e^{4A-\chi}\left(\frac{3}{2}\partial_{\rho} A + {\cal W}_2 \right)\right|_{\rho\rightarrow\rho_2}\,,
 \eeqs
which for ${\cal W}_2\,=\,-\frac{4}{3}\,-\,\frac{4}{3}\phi^2$, as can be verified directly by using the UV expansions in Eqs.~(\ref{eq:uvexpansion1})-(\ref{eq:uvexpansion4}), yields $ {\cal F}= -\frac{e^{4A^U-\chi^U}}{12}\left(4\phi_{\alpha}\phi_{\beta}\frac{}{}-\frac{}{}15\chi^{(5)}\right)=-\mu/4$.

\subsection{Scale setting and phase transition}
\label{Sec:scale}
 To enable a meaningful comparison of the free energy associated with different solutions, a scale-setting prescription is required. We implement this by introducing a common reference energy scale, $\Lambda$, defined as the inverse of the time needed for a massless particle to propagate from the ultraviolet region to the end of the geometry~\cite{Csaki:2000cx}:
\begin{equation}\label{eq:Lambda}
    \L^{-1}\equiv\int_{\r_0}^{\infty}\sqrt{\frac{g_{\r\r}}{|g_{tt}|}}d\r=\int_{\r_0}^{\infty}e^{\c(\r)-A(\r)}d\r=\int_{\vr_0}^{\infty}\frac{1}{\vr\sqrt{f(\vr)}}d\vr,
\end{equation}
This quantity is evaluated on the corresponding background geometry. For the main class of solutions under consideration for this work, the resulting integral does not admit a closed-form expression and must therefore be computed numerically. We then express all dimensional quantities, such as the free energy and the UV deformation parameters, in dimensionless form, by rescaling them with appropriate powers of $\Lambda$, as follows:
\beqs
\hat{\mathcal{F}}=\mathcal{F}\Lambda^{-5}\, ,\\
\hat{\phi}_\beta=\phi_\beta\Lambda^{-3}\, ,\\
\hat{A_6}^{(3),3}=A_6^{(3),3}\Lambda^{-3}\,, \\
\hat{\c}^{(5)}=\c^{(5)}\Lambda^{-5}\,. 
\eeqs

\begin{figure}[t]
\centering
\includegraphics[width=0.6\textwidth]{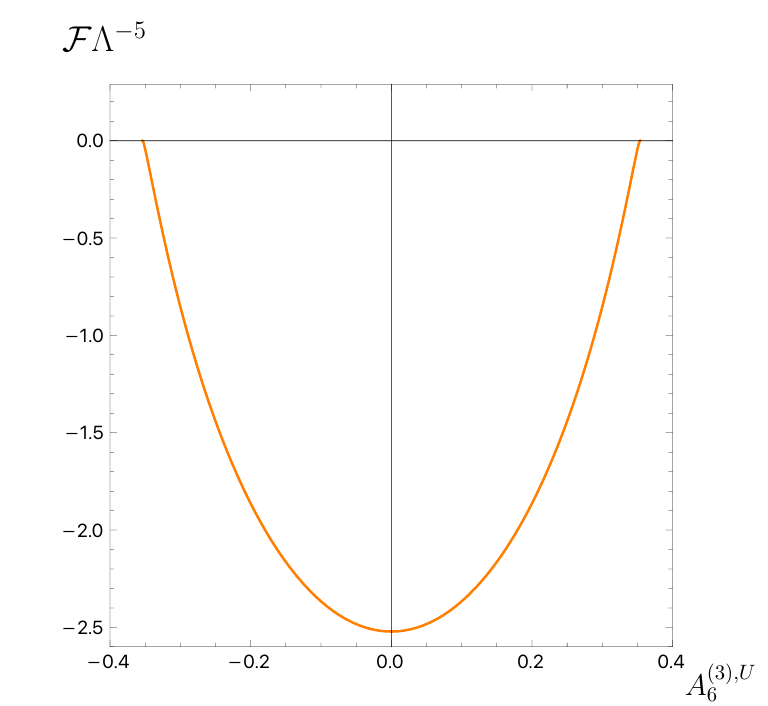}        
   \hfill
	\caption{The free energy density, $\hat{\cal F}={\cal F}\Lambda^{-5}$, expressed in units of $\Lambda$, associated to the soliton solutions with non-constant flux, $A_6^{(3)}$, that are dual to confining field theories, as a function of the deformation parameter,  $A^{(3),U}_6$. Domain-wall solutions with constant $A^{(3),U}_6$ exist for all values of $A^{(3),U}_6$. They are dual to conformal theories, and have ${\cal F}=0$. A first-order phase transition appears at $|A_6^{(3),U}|=A_6^{(3),U}({\rm max})=\frac{1}{2\sqrt{2}}\sim 0.35$.
  \label{fig:Freeenergy}}
\end{figure}

Figure~\ref{fig:Freeenergy} displays our results for the dimensionless free energy density, $\hat{\cal F}$,  computed for the solutions with non-trivial flux, as a function of the dimensionless deforming parameter $A_6^{(3),U}$. In the case of the AdS$_6$, domain-wall solutions with trivial (constant) flux, the prescription in Eq.~(\ref{Eq:ff}) yields ${\cal F}=0$, and these solutions exist for any choice of $A_6^{(3),U}$. The figure demonstrates the existence of a first-order phase transition taking place at the extremum of the branch of solutions with non-trivial flux: when $\mu=0$, there is coexistence of phases, with both the soliton and domain-wall solutions having vanishing free energy.
This observation allows to define a maximum value for $A_6^{(3),U}({\rm max})=\frac{1}{2\sqrt{2}}\sim 0.35$, obtained when $\vr_0=3/4$.  For  $|A_6^{(3),U}|>A_6^{(3),U}({\rm max})$, the AdS$_6$ solutions with constant $A_6^{(3)}$ are the only ones that exist, and have vanishing free energy. For $|A_6^{(3),U}|<A_6^{(3),U}({\rm max})$, both classes of solutions exist, but the soliton solutions with non-trivial $A_6^{(3)}$ have negative free energy, and hence they are dynamically favoured.

\section{Local stability analysis: fluctuations}
\label{Sec:spectra}

In this section we report the results of a local stability analysis, performed by applying the gauge-invariant formalism developed in Refs.~\cite{Bianchi:2003ug,Berg:2005pd,Berg:2006xy,Elander:2009bm, Elander:2010wd,Elander:2010wn, Elander:2014ola,Elander:2018aub,Elander:2020csd} to the fluctuations of fields retained in the gravity theory: the metric, the three scalars $\phi$, $\chi$, and $A^{(3)}_6$, and the two vectors $V_M$ and $A_M^{(3)}$. In the language of the dual field theory, they correspond to bound states with different (integer) spin. We perform this analysis only for the soliton solution with non-constant flux, for which an interpretation of the dual field theory in terms of confinement in four dimensions exists.

One starts by decomposing the fields in their five-dimensional form by splitting them in their background values and fluctuations. The metric is treated with the Arnowitt-Deser-Misner formalism (ADM)~\cite{Arnowitt:1959ah,Arnowitt:1962hi}, by writing
\beqs
\di s^2_5 &=& \left((1+\nu)^2\frac{}{}+\nu_{\mu}\nu^{\mu}\right)\di r^2 \,+\,2\nu_{\mu}\di x^{\mu}\di r \,+\, e^{2A(r)}\left(\eta_{\mu\nu}\frac{}{}+h_{\mu\nu}\right)\di x^{\mu}\di x^{\nu}\,,
\eeqs
where $\nu$, $\nu_{\mu}$,  and $h_{\mu\nu}$ are the fluctuations. The other fields are treated in analogy, by separating the background values, denoted with a $\bar{\quad}$, and their fluctuations, marked with a $\tilde{\quad}$, which we hence rewrite as follows:
\beqs
        \phi &=&  \bar \phi(r)  +\tilde{\phi} (x^{\mu},\,r)= \phi(r) +\tilde{\phi}(x^{\mu},\,r)\\
        		\c &=&  \bar \c(r)  +\tilde{\c}(x^{\mu},\,r) = \c(r) +\tilde{\c}(x^{\mu},\,r)\,,\\
		A_6^{(3)}&=& \bar{A}^{(3)}_6(r)  +\tilde{A}^{(3)}_6(x^{\mu},\,r) = A_6^{(3)}(r) +\tilde{A}^{(3)}_6(x^{\mu},\,r)\,,\\
	V_M &=&  \bar V_M(r)  + \tilde{V}_M(x^{\mu},\,r) = \tilde{V}_M(x^{\mu},\,r)\,,\\
		A_M^{(3)} &=&   \bar A_M^{(3)}(r)  +\tilde{A}_M^{(3)}(x^{\mu},\,r) = \tilde{A}_M^{(3)}(x^{\mu},\,r)
	\,.
\eeqs
We made explicit the fact that the background functions depend only on the holographic direction in the geometry, while the fluctuations depend also on the four-dimensional space-time coordinates, $x^{\mu}$.

\begin{figure}[t]
\centering
\includegraphics[width=0.6\textwidth]{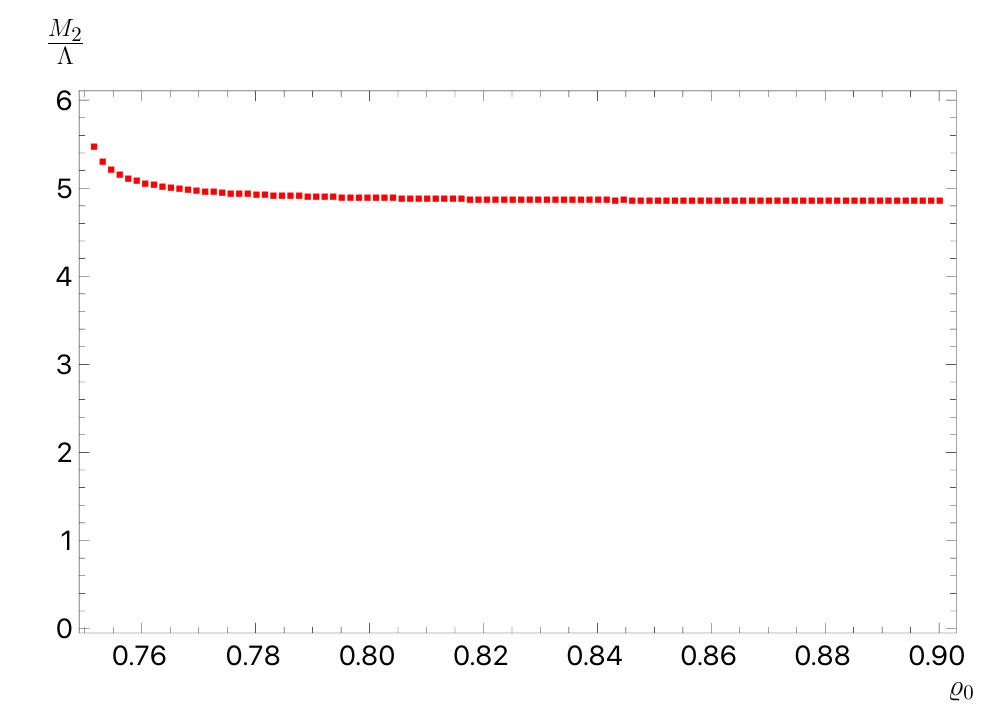}        
   \hfill
	\caption{Ratio, $M_2/\Lambda$, of the mass of the lightest spin-2 bound state of the confining field theory dual to the soliton solutions with non-trivial $A_6^{(3)}$, over the scale $\Lambda$ defined in Eq.~(\ref{eq:Lambda}). The mass spectrum has been computed by applying the midpoint determinant method to the linearised, gauge-invariant equations obeyed by the tensor fluctuations, $\mathfrak{e}_\n^\m$, as a function of the end-of-space parameter, $\vr_0$. In the numerical analysis, Neumann boundary conditions have been imposed at   $\vr_1=\vr_0+ 10^{-7}$, and $\vr_2=\vr_0+10$. We verified that with these choices the spectrum is close enough to the physical limit that the IR ($\vr_1\rightarrow \vr_0$) and UV ($\vr_2\rightarrow +\infty$) regulators have no discernible effect, given our numerical precision goals.
  \label{fig:ScaleLambda}}
\end{figure}

\subsection{Tensor spectrum}
\label{Sec:spin2}
As described in Appendix~\ref{Sec:gaugeinvariantformalism}, the calculation of the tensor fluctuations, that correspond to spin-2 bound states of the dual field theory, requires first  to  decompose the fluctuation $h_{\mu\nu}$ as follows:
\beqs
h^{\mu}_{\,\,\,\nu}&=&\mathfrak{e}^{\mu}_{\,\,\,\nu}\,+iq^{\mu}\epsilon_{\nu}\,+i\epsilon^{\mu}q_{\nu}\,+\frac{q^{\mu}q_{\nu}}{q^2}H\,+\frac{1}{3}\delta^{\mu}_{\,\,\,\nu}h\,.
\eeqs
We focus our attention on the spin-2  fluctuations, that are encoded in the traceless, symmetric, gauge-invariant fields, $\mathfrak{e}^{\mu}_{\,\,\,\nu}$, that obey by the following linear equations of motion:
\beqs\label{eq:tensorEoM}
\left[\6_r^2 +4\6_r A\6_r +e^{-2A}M^2\right]\mathfrak{e}_\n^\m=0\,.
\eeqs
We can rewrite these equations, after the  change of variable $\frac{\partial \rho}{\partial r} = e^{-\chi}$, as
\beqs
\left[\6_\r^2 +(4\6_\r A-\6_\r\c)\6_\r+e^{2(\c-A)}M^2 \right]\mathfrak{e}_\n^\m=0\,,
\eeqs
and again, after the change of variable $\frac{\partial \vr}{\partial \rho} = \frac{1}{\vr^2}e^{2A+\chi}=\frac{\sqrt{f}}{H^{1/4}}$, in the form
\beqs
\left[\vr \6_\vr^2 +(6 \vr \6_\vr A-2)\6_\vr+e^{-6A}\vr^5 M^2 \right]\mathfrak{e}_\n^\m=0\,,
\eeqs
or, equivalently, as
\beqs\label{eq:tensorEoMH}
\left[\vr \6_\vr^2 +\left(\frac{\vr f'(\vr)}{f(\vr)}+4\right)\6_\vr+\frac{ M^2}{\vr f(\vr)} \right]\mathfrak{e}_\n^\m=0\,.
\eeqs
The fluctuations are subject to Neumann boundary condition, which, by performing the same changes of variable can be written in any of the following forms:
\beqs
\label{Eq:bcT}
\left.\frac{}{}\6_r\mathfrak{e}_\n^\m\right|_{r_i}
=\left.\frac{}{}e^{-\c}\6_\r\mathfrak{e}_\n^\m\right|_{\r_i}
=\left.\frac{}{}\frac{e^{2A}}{\vr^2}\6_\vr\mathfrak{e}_\n^\m\right|_{\vr_i}
=\left.\frac{}{}f^{1/3}(\vr)\6_\r\mathfrak{e}_\n^\m\right|_{\vr_i}
=0\,.
\eeqs

As we write the background functions in closed form in the variable $\vr$, we find it convenient to adopt this same choice of coordinate also for the fluctuations. We parameterise the family of solutions in terms of the position of the end of space in this variable, which is constrained to lie inside the interval $\frac{3}{4} <\vr_0<\frac{9}{10}$. In performing the numerical calculations, we adopt the mid-point determinant, as in Ref.~\cite{Berg:2006xy}. We scan numerically over the values of $M^2$, in the presence of  explicit cutoffs $\vr_0<\vr_1<\vr<\vr_2<+\infty$, by imposing the boundary conditions in Eq.~(\ref{Eq:bcT}) at the two cutoffs, evolving the solutions of Eq.~(\ref{eq:tensorEoMH}) to an intermediate value of the radial direction, $\vr^{\ast}$, computing the determinant of the $2 \times 2$ matrix formed with the values of $\mathfrak{e}_\n^\m$ and its derivative, evaluated at $\vr^{\ast}$, for both solutions coming from the UV and IR, and identifying the zeros of the determinant by varying $M^2$.  When the determinant vanishes, this signals the fact that the evolutions of the linearised equations from the IR ($\vr_1$) and from the UV ($\vr_2$) yield coinciding fluctuations, up to an unphysical multiplicative normalisation, and hence these values of $M^2$ yield the mass spectrum of bound states of interest. We checked that for asymptotic choices of the regulators, $\vr_{1,2}$, the resulting spectrum converges to the physical results, by verifying that there are no residual dependencies on $\vr_1$ or $\vr_2$. We find that the choices $\vr_1=\vr_0+ 10^{-7}$, and $\vr_2=\vr_0+10$ are sufficient to achieve these goals, with the level of numerical precision we require in this study.

In Fig.~\ref{fig:ScaleLambda}, we show the mass, $M_2$, of the lightest tensor fluctuation, normalised to the scale, $\Lambda$, defined in Eq.~(\ref{eq:Lambda}), as a function of the parameter, $\vr_0$, that labels the end-of-space point of the solutions dual to confining field theories. The ratio of these two quantities is approximately constant, except for the region in close proximity to the end of the branch of solution, $\vr_0\simeq 3/4$, in which case the discrepancy is still a modest $(10 - 20)\,\%$. This observation demonstrates that these two scales are closely related to one another, and can be used interchangeably at representative of the confinement scale of the theory. While we used $\Lambda$ in the study of the free energy, we elect to use $M_2$ in the study of the spectrum of fluctuations presented in later subsections of the paper.

\subsection{Scalar spectrum}
\label{Sec:spin0}

Following again Appendix~\ref{Sec:gaugeinvariantformalism}, in $D=5$ dimensions, the three gauge-invariant scalar fluctuations, which correspond to combinations of the fluctuations of the trace of the metric, $h$, with those of the three scalar fields, $\tilde{\phi}$, $\tilde{\chi}$, and $\tilde{A}_6^{(3)}$, satisfy the following system of coupled, linear second-order equations 
\begin{equation}\label{eq:ScalarEoM}
	\begin{split}
	0 &= \left[ \mathcal{D}^2_r + 4\partial_r A \mathcal{D}_r  + e^{-2A}M^2 \right] \mathfrak{a}^a + \\
	 &- \left[ V^a _{\,\,\,|c} - R_{\,\,\,bcd}^a \partial_r \bar{\Phi}^b \partial_r \bar{\Phi}^d + \frac{4 (\partial_r \bar{\Phi}^a V^b + V^a \partial_r \bar{\Phi}^b)G_{bc}}{3 \partial_r A} + \frac{ 16 V \partial_r \bar{\Phi}^a \partial_r \bar{\Phi}^b G_{bc}}{9 (\partial_r A)^2} \right] \mathfrak{a}^c\,,
	\end{split}
\end{equation}
where $\bar{\Phi}^a\,=\,\left\{\bar{\phi},\,\bar{\chi},\,\bar{A}_6^{(3)}\right\}$ are the background scalar fields. The fluctuations are subject to the boundary condition:
\begin{equation}
    \6_r\phi^a\6_r\phi_b\mathcal{D}_r\mathfrak{a}^b\Big|_{r_i}=\left[-\frac{3}{2}\6_rAe^{-2A}M^2\delta_b^a+\6_r\phi^a\left(\frac{4V}{3\6_rA}\6_r\phi_b+V_b\right)\right]\mathfrak{a}^b\Big|_{r_i}\,,
\end{equation}
where  $\mathcal{D}_r\mathfrak{a}^a\equiv\partial_r\mathfrak{a}^a +\mathcal{G}^a_{bc}\partial_r\Phi^b\mathfrak{a}^c$, while  	$V^a_{\,\,\,|b}\equiv\partial_b V^a + \mathcal{G}^a_{\,\,\,bc}V^c$, and	$V^a\equiv G^{ab}\partial_bV$.

\begin{figure}[t]
\centering
			{\includegraphics[width=0.6\textwidth]{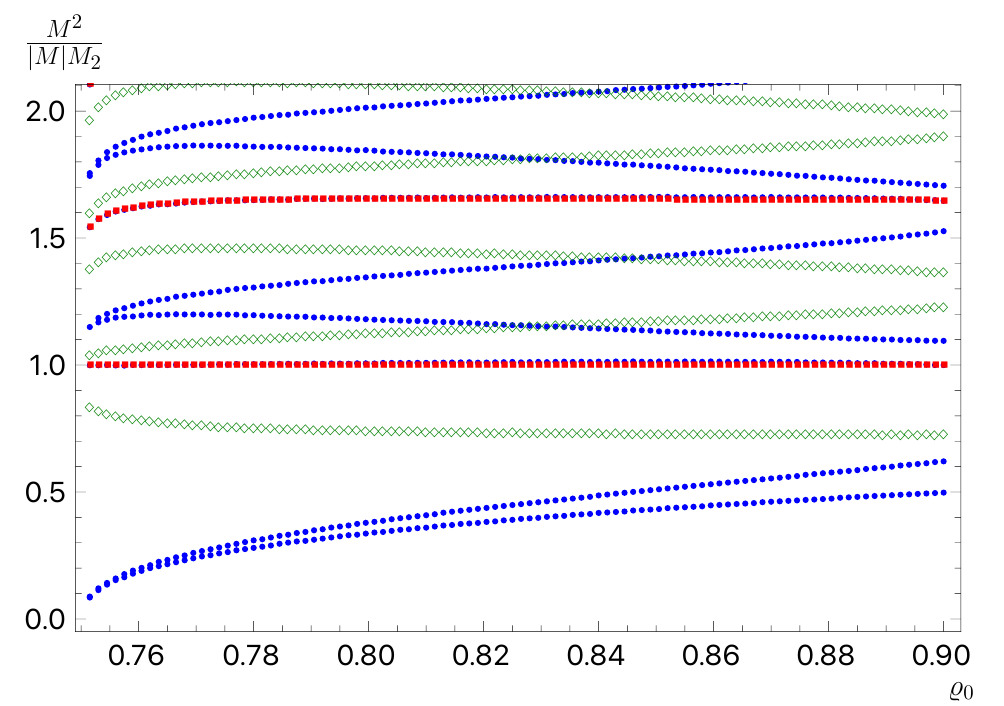}}
    \caption{Spectrum of masses, $M^2/|M|$, of the fluctuations of the confining backgrounds, expressed in units of the mass of the lightest tensor in the spectrum, $M_2$.  The tensor states, coming from (symmetric, transverse, and traceless)  fluctuations of the metric, are shown in red. The scalar states, arising as the three gauge invariant combinations of the fluctuations of the three sigma-model scalars, combined with the trace of the metric, are in blue. The gauge-invariant, transverse fluctuations of the two vectors are depicted in green. The numerical solution of the fluctuation equations are performed using IR and UV cutoffs set to $\vr_1=\vr_0+10^{-7}$ and $\vr_2=\vr_0+10$, respectively. 
 }
    \label{fig:mass_spectrum}
\end{figure}

For the numerical calculations of the spectrum, we found that  imposing the boundary conditions directly on the numerical solutions of the differential equations results in the slow convergence of the results towards the physical limit. To improve convergence, we deployed a different process, with respect to the case of the tensor fluctuations. First, we perform expansions of the IR and UV asymptotic behaviour of the solutions, which we report in Appendix~\ref{Sec:UVfluc}. We impose the boundary conditions on the expansions, hence removing the dominant contributions and retaining only the sub-leading ones. We then introduce the same cutoffs as for the tensor modes, $\vr_1$ and $\vr_2$, as regulators. But the boundary conditions we impose on the numerical solutions are dictated by matching them to the constrained UV and IR expansions obtained by retaining only the subleading terms. We then apply the mid-determinant method to identify the physical spectrum, $M^2=-q^2$, having evolved the numerical solutions towards a mid point, $\vr^{\ast}$. Our numerical results for the spectrum of gauge-invariant  fluctuations are provided in Fig.~\ref{fig:mass_spectrum}, which combines the spin-0 (scalar), spin-2 (tensor), and  spin-1 (vector) modes. We will comment on the results later in the paper, after we explain how we computed the spectrum of vector modes.

\subsection{Vector spectrum}
\label{Sec:spin1}

The study of vector fluctuations, dual to spin-1 bound states in the field theory, is more subtle, as it involves internal gauge symmetries (in addition to the diffeomorphisms), and some non-trivial effect due to mixing between vectors induced by the non-vanishing of $A_6^{(3)}$. Hence, we provide here a more detailed discussion of the process of deriving the equations of motion and boundary conditions for the transverse component of the gauge fields. 

 Expanding the last term of the action, Eq.~(\ref{Eq:S5astprime}), that encodes the vector fields, by substituting the definitions of the fluctuations, but retaining only terms quadratic in the fluctuations, one arrives to the following:
\begin{widetext}
\beqs
\mathcal{S}_V = \int \dd^5x \sqrt{-g_5}\,\,g^{MR}g^{NS}&&\Bigg[-\frac{1}{16}e^{8\c}\tilde{F}_{MN}\tilde{F}_{RS}-\frac{1}{4}e^{2\c-2\phi}\tilde{A}^{(3)}_{MN}\tilde{A}^{(3)}_{RS}
\\
\nonumber
&&-\frac{1}{4}e^{2\c-2\phi}\left(\tilde{V}_M\6_N A^{(3)}_6-\tilde{V}_N\6_M A^{(3)}_6\right)\left(\tilde{V}_R\6_S A^{(3)}_6-\tilde{V}_S\6_R A^{(3)}_6\right)
\\
\nonumber
&&-\frac{1}{2}e^{2\c-2\phi}\tilde{A}_{MN}^{{(3)}}\left(\tilde{V}_R\partial_SA_6^{(3)}-\tilde{V}_S\partial_RA_6^{(3)}\right)\Bigg]\,,
\nonumber
\eeqs
\end{widetext}
where $\tilde{F}_{MN}=\partial_M\tilde{V}_N-\partial_N\tilde{V}_M=F^V_{MN}$, while  $\tilde{A}^{(3)}_{MN}=\partial_M\tilde{A}^{(3)}_N-\partial_N\tilde{A}^{(3)}_M$ is only the non-interacting portion of $F_{MN}^{(3)}$.

In order to show explicitly how gauge invariance works in this case, it is useful to write explicitly the decomposition of the five-dimensional vector fields in their four-dimensional vector and  scalar components. To do so, we use the compact notation $\tilde{V}^A_M=\{\tilde{V}_M,\,\tilde{A}^{(3)}_M\}$ and $H_{AB}={\rm diag} \left(\frac{1}{4}e^{8\chi},\,e^{2\chi-2\phi}\right)$, use the ansatx for the five-dimensional metric, perform a Fourier transformation in the $x^{\mu}$ coordinates, that introduces the momentum $q^{\mu}$ in place of the derivatives, and, after some integrations by parts, write
\beqs\label{eq:FourierV}
		\mathcal{S}_{V}&=&\int \dd^4q \dd r  \left\{H_{AB}\left[-\frac{1}{2}\tilde{V}_{\m}^A(-q)q^2P^{\m\n}\tilde{V}_{\n}^B(q)-\frac{1}{2}e^{2A(r)}q^2\tilde{V}_5^A(-q)\tilde{V}_5^B(q)\right]\right.\\ 
\nonumber		&&+\frac{1}{2}\tilde{V}_{\m}^A(-q)\eta^{\m\n}\left[\partial_r(H_{AB}e^{2A(r)}\partial_r\tilde{V}_{\n}^B(q))\right]\\
\nonumber			&&+\sum_{i=1, 2}(-)^i\delta(r-r_i)\left[-\frac{1}{2}H_{AB}e^{2A(r)}\tilde{V}_{\m}^A(-q)\eta^{\m\n}\partial_r\tilde{V}_{\n}^B(q)\right]\\
\nonumber			&&-\frac{1}{2}\left[iq^{\m}\tilde{V}_{\m}^A(-q)\partial_r(H_{AB}e^{2A(r)}\tilde{V}_5^B(q))+(q\leftrightarrow-q)\right]\\
\nonumber			&&+\sum_{i=1,2}(-)^i\d(r-r_i)\left[\frac{i}{2}q^{\m}\tilde{V}_{\m}^A(-q)(H_{AB}e^{2A(r)}\tilde{V}_5^B(q))+(q\leftrightarrow-q)\right]\\
\nonumber			&&-\frac{1}{2}e^{2A(r)}e^{2\c-2\phi}\eta^{\m\n}\tilde{V}_{\m}(-q)\tilde{V}_{\n}(q)(\partial_rA_6^{{(3)}})^2\\\nonumber			&&\left.\frac{}{}-\frac{1}{2}e^{2A(r)}\partial_r A^{(3)}_6e^{2\c-2\phi}\eta^{\m\n}\left(iq_{\n}\tilde{V}_{\m}(-q)\tilde{A}^{(3)}_5(q)-\tilde{V}_{\m}(-q)\partial_r\tilde{A}^{(3)}_\n(q)+(q\leftrightarrow -q)\right)\right\}\,.
\nonumber
\eeqs
Here, $P^{\m\n}\equiv \eta^{\mu\nu}-\frac{q^{\mu}q^{\nu}}{q^2}$ is the transverse momentum projector, that satisfies the relation $P^{\m\n}q_{\nu}=0$. The boundary terms are remnants of the integrations by parts.

Recalling that the matrix $H_{AB}$ is diagonal, we notice that the third, fourth and fifth terms in the right-hand side of Eq.~(\ref{eq:FourierV}) are the familiar ones, appearing in the Kaluza-Kline decomposition of a vector, because we are considering the holographic direction as separate from the Minkowski directions. In addition, we observe the presence of mixing terms between the vector field that arose as a gravi-photon from the circle compactification, $\tilde{V}_{\m}$,  and the (pseudo-)scalar associated with the Abelian gauge field, $\tilde{A}^{(3)}_5$, as can be seen in the last term of  Eq.~(\ref{eq:FourierV}).  Loosely speaking, $\tilde{A}^{(3)}_5$ is the field component that is Higgsed into $\tilde{V}_{\m}$ due to the emergence of a non-trivial background value of ${A}^{(3)}_6$. The mixing terms between spin-1 and spin-0 field components in the bulk of the geometry can be eliminated by supplementing the bulk action with the following gauge-fixing terms:
\beqs
		\mathcal{S}_{\x}&=&\int \dd^4q \dd r  \sum_{A=V,A^{(3)}}\nonumber
		\left[-\frac{H_{AA}}{2\x}\left(q^{\n}\tilde{V}_{\n}^A(-q)
       +i\frac{\x}{H_{AA}}\partial_r(H_{AA}e^{2A(r)}\tilde{V}_5^A(-q))
       +i\frac{\xi}{H_{AA}}\d_A^V e^{2A(r)}\6_rA^{(3)}_6e^{2\c-2\phi}\tilde{A}^{(3)}_5(-q)\right)
		\right.\\
	&&	      \times\left.
		\left(q^{\m}\tilde{V}_{\m}^A(q)
		-i\frac{\x}{H_{AA}}\partial_r\left(H_{AA}e^{2A(r)}\tilde{V}_5^A(q)\right)
		-i\frac{\xi}{H_{AA}}\d^V_A e^{2A(r)}\6_rA^{(3)}_6e^{2\c-2\phi}\tilde{A}^{(3)}_5(q)\right)
 \right]\,.
\eeqs

\allowdisplaybreaks
By adding these gauge fixing terms, the action for the vector sector is:
\beqs \label{eq:gauged-fixed-action-1}
	\mathcal{S}_{V} +\mathcal{S}_{\x}&=&\int \dd^4q \dd r 
	\left\{ -\frac{H_{AB}}{2}\tilde{V}_{\m}^A(-q)q^2\left(P^{\m\n}+\frac{1}{\x}\frac{q^\m q^\n}{q^2}\right)\tilde{V}_{\n}^B(q)\right.\\
    \nonumber    &&+\frac{1}{2}\tilde{V}_{\m}^A(-q)\left(P^{\m\n}+\frac{q^\m q^\n}{q^2}\right)\partial_r\left(H_{AB}e^{2A(r)}\partial_r\tilde{V}_{\n}^B(q)\right)\\
	\nonumber	&&{-}\frac{1}{2}e^{2A}e^{2\c-2\phi}\left(\partial_r A^{(3)}_6\right)^2\tilde{V}_{\m}(-q)\left(P^{\m\n}+\frac{q^\m q^\n}{q^2}\right)\tilde{V}_{\n}(q)\\
	\nonumber	&&+\frac{1}{2}e^{2A}\partial_r A^{(3)}_6e^{2\c-2\phi}\left[\tilde{V}_{\m}(-q)\left(P^{\m\n}+\frac{q^\m q^\n}{q^2}\right)\partial_r\tilde{A}^{(3)}_{\n}(q)+(q \leftrightarrow -q)\right]\\
	\nonumber	&& -\frac{1}{2}e^{2A(r)}H_{AB}q^2\tilde{V}_5^A(-q)\tilde{V}_5^B(q)\\
  \nonumber      &&-\sum_{A=V,A^{(3)}}\frac{\xi}{2H_{AA}}\6_r\left(H_{AA}e^{2A}\tilde{V}^A_5(-q)\right)\6_r\left(H_{AA}e^{2A}\tilde{V}^A_5(q)\right)\\
    \nonumber    &&-\frac{\xi}{2}\left(e^{2A}\6_rA^{(3)}_6\right)^2\frac{e^{2\c-2\phi}}{H_{VV}}\tilde{A}^{(3)}_5(-q)\tilde{A}^{(3)}_5(q)\\
     \nonumber   &&+\frac{\xi}{2} e^{2A}\frac{e^{2\c-2\phi}}{H_{VV}}\6_rA^{(3)}_6\left(\6_r(H_{VV}e^{2A}\tilde{V}_5(-q))\tilde{A}^{(3)}_5(q)-(q\leftrightarrow-q)\right)\\
        	\nonumber&&+\sum_{i=1, 2}(-)^i\delta(r-r_i)\left[-\frac{1}{2}H_{AB}e^{2A(r)}\tilde{V}_{\m}^A(-q)\left(P^{\m\n}+\frac{q^\m q^\n}{q^2}\right)\partial_r\tilde{V}_{\n}^B(q)\right]\\
		&&\left.\frac{}{}+\sum_{i=1,2}(-)^i\d(r-r_i)\left[\frac{i}{2}q^{\m}\tilde{V}_{\m}^A(-q)(H_{AB}e^{2A(r)}\tilde{V}_5^B(q))+(q\leftrightarrow-q)\right]\nonumber
		\right\}\,.
\eeqs

The last term in Eq.~(\ref{eq:gauged-fixed-action-1}) requires to  complement the bulk gauge-fixing action with  boundary-localised gauge fixing terms:
\beqs
    \mathcal{S}_M=\int \dd^4q \dd r &&\sum_{i=1, 2}(-1)^i\d(r-r_i)\sum_{A=V,A^{(3)}}\left[-\frac{1}{2M_i}\left(q^\m \tilde{V}_\m^A(-q)-iM_iH_{AA}e^{2A(r)}\tilde{V}_5^A(-q)\right)\times\right.\\
    &&\left.\frac{}{}\times \left(q^\n\tilde{V}_\n^A(q)+iM_iH_{AA}e^{2A(r)}\tilde{V}_5^A(q)\right)\right]\,.\nonumber
\eeqs

Finally, the gauge-fixed action for the spin-1 sector is the following:
\beqs \label{eq:gauged-fixed-action-2}
	\mathcal{S}_{V} +\mathcal{S}_{\x} +\mathcal{S}_{M}&=&\int \dd^4q \dd r 
	\left\{ -\frac{H_{AB}}{2}\tilde{V}_{\m}^A(-q)q^2\left(P^{\m\n}+\frac{1}{\x}\frac{q^\m q^\n}{q^2}\right)\tilde{V}_{\n}^B(q)\right.\\
    \nonumber    &&+\frac{1}{2}\tilde{V}_{\m}^A(-q)\left(P^{\m\n}+\frac{q^\m q^\n}{q^2}\right)\partial_r\left(H_{AB}e^{2A(r)}\partial_r\tilde{V}_{\n}^B(q)\right)\\
	\nonumber	&&{-}\frac{1}{2}e^{2A}e^{2\c-2\phi}\left(\partial_r A^{(3)}_6\right)^2\tilde{V}_{\m}(-q)\left(P^{\m\n}+\frac{q^\m q^\n}{q^2}\right)\tilde{V}_{\n}(q)\\
	\nonumber	&&+\frac{1}{2}e^{2A}\partial_r A^{(3)}_6e^{2\c-2\phi}\left[\tilde{V}_{\m}(-q)\left(P^{\m\n}+\frac{q^\m q^\n}{q^2}\right)\partial_r\tilde{A}^{(3)}_{\n}(q)+(q \leftrightarrow -q)\right]\\
	\nonumber	&& -\frac{1}{2}e^{2A(r)}H_{AB}q^2\tilde{V}_5^A(-q)\tilde{V}_5^B(q)\\
  \nonumber      &&-\sum_{A=V,A^{(3)}}\frac{\xi}{2H_{AA}}\6_r\left(H_{AA}e^{2A}\tilde{V}^A_5(-q)\right)\6_r\left(H_{AA}e^{2A}\tilde{V}^A_5(q)\right)\\
    \nonumber    &&-\frac{\xi}{2}\left(e^{2A}\6_rA^{(3)}_6\right)^2\frac{e^{2\c-2\phi}}{H_{VV}}\tilde{A}^{(3)}_5(-q)\tilde{A}^{(3)}_5(q)\\
     \nonumber   &&+\frac{\xi}{2} e^{2A}\frac{e^{2\c-2\phi}}{H_{VV}}\6_rA^{(3)}_6\left(\6_r(H_{VV}e^{2A}\tilde{V}_5(-q))\tilde{A}^{(3)}_5(q)-(q\leftrightarrow-q)\right)\\
        	\nonumber&&+\sum_{i=1, 2}(-)^i\delta(r-r_i)\left[-\frac{1}{2}H_{AB}e^{2A(r)}\tilde{V}_{\m}^A(-q)\left(P^{\m\n}+\frac{q^\m q^\n}{q^2}\right)\partial_r\tilde{V}_{\n}^B(q)\right]\\
				&&\left.\frac{}{}+\sum_{i=1,2}(-)^i\d(r-r_i)\left[
		-\frac{1}{2M_i}(q^{\m}q^{\n}\d_{AB}\tilde{V}_{\mu}^A(-q)\tilde{V}_{\nu}^B(q))
		-\frac{M_i}{2}(H_{AB})^2e^{4A(r)}\tilde{V}^A_5(-q)\tilde{V}^B_5(q)
		\right]\nonumber
		\right\}\,.
\eeqs

The physical spectrum for spin-1 states is obtained by focusing on the transverse part of the vector fields, containing  $P^{\mu\nu}V_{\nu}^A$, while ignoring terms containing $\tilde{V}^A_5$ and $q^{\mu}q^{\nu}\tilde{V}^A_{\nu}$, that are now decoupled from the former. By taking variations in respect to $\tilde{A}_{\mu}^{(3)}$, we find that the pertinent classical equations of motion and boundary conditions  are
\beqs
0&=&\left[-\frac{1}{2}e^{2\c-2\phi}q^2+\frac{1}{2}\6_r\left(e^{2\c-2\phi+2A}\6_r\right)\right]P^{\m\n}\tilde{A}_{\n}(q, r)
-\frac{1}{2}\6_r\left[\left(e^{2\c-2\phi+2A}\6_rA^{(3)}_6\right)P^{\m\n}\tilde{V}_{\n}(q, r)\right]\,,\\
0&=&\left.\frac{}{}\left[-\frac{1}{2}e^{2\c-2\phi+2A}\6_r\right]P^{\m\n}\tilde{A}_{\n}(q)\right|_{r=r_i}+\left.\frac{}{}\left[\frac{1}{2}e^{2\c-2\phi+2A}\6_rA^{(3)}_6\right]P^{\m\n}\tilde{V}_{\n}(q)\right|_{r=r_i}\,.
\eeqs
By taking variations in $\tilde{V}_{\mu}$, we arrive at 
\beqs
0&=&\left[-\frac{1}{8}e^{8\c}q^2+\frac{1}{8}\6_r\left(e^{8\c+2A}\6_r\right){-}\frac{1}{2}e^{2\c-2\phi+2A}\left(\6_rA^{(3)}_6\right)^2\right]P^{\m\n}\tilde{V}_{\n}(q, r)\\
&&\nonumber+\left[\frac{1}{2}e^{2\c-2\phi+2A}\6_rA^{(3)}_6\6_r\right]P^{\m\n}\tilde{A}_{\n}(q, r)\,,\\
0&=&\left.\frac{}{}\left[-\frac{1}{8}e^{8\c+2A}\6_r\right]P^{\m\n}\tilde{V}_{\n}(q, r)\right|_{r=r_i}\,.
\eeqs

Upon making the replacement $q^2 = -M^2$, and simplifying common terms, the physical equations of motion and the corresponding boundary conditions for the vector fluctuations take the final form:
\beqs
  0&=& \left[e^{-2A}M^2+\frac{}{}2\left(\partial_r\c-\partial_r\phi+\partial_rA\right)\6_r+\6_r^2\right]P^{\m\n}\tilde{A}_{\n} \\
&&\nonumber  -\left[2\left(\partial_r\c-\partial_r\phi+\partial_rA\right)\left(\6_rA^{(3)}_6\right)+\6_r^2A^{(3)}_6+\left(\6_rA^{(3)}_6\right)\6_r\right]P^{\m\n}\tilde{V}_{\n}\,,\\
 0&=& \left.\frac{}{} -\6_r\left[P^{\m\n}\tilde{A}_{\n}(q)\right]\right|_{r=r_i}+\left.\frac{}{}\left[\6_rA^{(3)}_6\right]P^{\m\n}\tilde{V}_{\n}(q)\right|_{r=r_i}\,,\\
  0&=& \left[e^{-2A}M^2{-}4e^{-6\c-2\phi}\left(\6_rA^{(3)}_6\right)^2+\left(8\partial_r\c+2\partial_r A\right)\6_r+\6_r^2\right]P^{\m\n}\tilde{V}_{\n}+\left[4e^{-6\c-2\phi}\left(\6_rA^{(3)}_6\right)\6_r\right]P^{\m\n}\tilde{A}_{\n}\,,\\
0&=&  \left.\frac{}{} \6_r\left[P^{\m\n}\tilde{V}_{\n}(q, r)\right]\right|_{r=r_i}\,.
\eeqs

As for the scalars, the procedure we apply in our numerical calculations of the spectrum contains an additional step, with respect to the case of the tensors. We introduce the same cutoffs, $\vr_1$ and $\vr_2$, as regulators, but rather than imposing the boundary conditions on the fields at these finite cutoffs, we perform expansions of the IR and UV asymptotic behaviour of the solutions, which we report in Appendix~\ref{Sec:UVfluc}, and impose the boundary conditions on said expansions, which amounts to retaining only the sub-leading terms. We then use the values of the expansions, evaluated at $\vr_1$ and $\vr_2$, to set up the numerical evolution of the linearised equations, and  apply again the mid-determinant method to identify the physical spectrum, $M^2$.

\subsection{Discussion and probe approximation}
\label{Sec:probe}

\begin{figure}[t]
\centering
			{\includegraphics[width=0.6\textwidth]{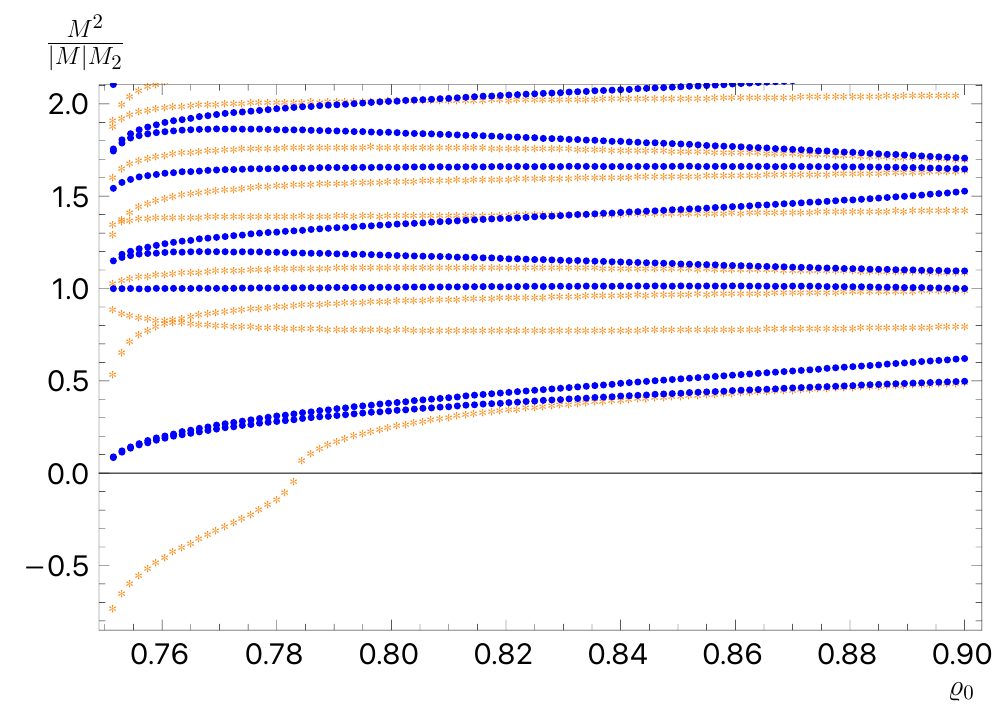}}
 \caption{Spectrum of masses, $M^2/|M|$, of the gauge-invariant scalar fluctuations of the soliton backgrounds with non-trivial flux (dual to confining field theories) expressed in units of the mass of the lightest tensor in the spectrum, $M_2$.  The scalar states, arising as the three gauge invariant combinations of the fluctuations of the three scalars together with the trace of the metric, are in blue. We compare the results with those of applying the probe approximation to the same states, depicted in orange, This approximation consists of explicitly neglecting the mixing of the fluctuation of the background scalars with the trace of the metric, hence ignoring the effect of the dilaton operator in the field theory. The numerical solution of the fluctuation equations is performed using IR and UV cutoffs set to $\vr_1=\vr_0+10^{-7}$ and $\vr_2=\vr_0+10$, respectively. 
 }
    \label{fig:mass_spectrumProbe}
\end{figure}

The spectrum of gauge-invariant fluctuations of the soliton backgrounds with non-trivial flux, which correspond to physical bound states in the dual field theory,  is provided in Fig.~\ref{fig:mass_spectrum}. The masses are shown as a function of $\vr_0$, over the whole parameter space. We normalise the masses of the bound states to that of the lightest spin-2 excitation, $M_2$, which is a good proxy for the confinement scale, as discussed earlier in the paper. The general appearance of the spectrum of heavy modes is the one expected from supergravity; for generic values of the parameter $\vr_0$, and for masses larger than $M_2$, we observe the presence of six towers of states, each tower comprising excitations that have a common periodicity, ${\cal O}(0.6\,M_2)$. The six towers correspond to the fluctuations of the three scalars, two vectors, and one tensor retained in the truncation. We note that one of the scalars is approximately degenerate with the tensor, making it difficult to render graphically in the figure.

There are several interesting features appearing in the lightest excitations. For generic values of $\vr_0$, we find a near-degenerate pair of spin-0 states to be the lightest states in the spectrum. One of the spin-1 states is also lighter than $M_2$. Interestingly, when approaching the lower limit of the parameter space, $\vr_0\rightarrow \frac{3}{4}$, several qualitatively new features appear. Firstly, the mass of the two lightest scalars is suppressed, in respect to $M_2$, and appears to vanish in the limit $\vr_0\rightarrow \frac{3}{4}$. Second, two of the scalar towers become approximately degenerate. This might be related to symmetry enhancement, but unfortunately the limiting case $\vr_0=\frac{3}{4}$ is singular. Understanding the features of the spectrum in this region of parameter space may require going beyond the supergravity approximation. We hence defer further commenting on the range of $\vr_0$ in close proximity to its lower extremum, except for observing that a similar behaviour has been observed also near one the two critical points of the theory discussed in Ref.~\cite{Elander:2025fpk} (see bottom-left corner of Fig.~6  in that publication).

For generic values of $\vr_0$, the two lightest scalars have mass in the range $(0.3\div 0.6)M_2$, suppressed, but not parametrically, in respect to the rest of the spectrum. It is interesting to investigate the nature of these two scalars. We do so by repeating the calculation of the spectrum of scalars using the probe approximation, as defined in Ref.~\cite{Elander:2020csd}. The starting point of the analysis is the observation that the gauge-invariant scalar fluctuation, $\mathfrak{a}^a$, defined in Eq.~(\ref{eq:scalarfluc}), contains a component along the trace of the metric, $h$, and that the coefficient controlling the mixing effects is proportional to the derivative of the scalar field evaluated on the background. Hence, for scalar fluctuations related to scalars that do not affect the background geometry via their background profile, this term can be neglected. In the dual language of field theory, $h$ is related to the trace of the stress-energy tensor, or the dilatation operator. The probe approximation consists of explicitly neglecting the contribution of $h$ to $\mathfrak{a}^a$, hence assuming that the dilatation operator cannot source the corresponding state.

We report in Appendix~\ref{Sec:probeEquations} the general form of the equations for the scalars in probe approximation, lifted from Ref.~\cite{Elander:2020csd}. We write explicitly the form of the equations in the backgrounds of interest in Appendix~\ref{Sec:eom}. We performed the numerical calculation of the spectrum in the probe approximation using the improvement given by 
imposing the boundary conditions on the UV and IR expansions of the fluctuations, and matching them with the numerical solutions of the differential equations evaluated at the cutoffs $\vr_1$ and $\vr_2$, along the lines of the process adopted for the gauge-invariant scalar and vector fluctuations. We report the form of the relevant expansions in Appendix~\ref{Sec:UVfluc}.

We show our results in Fig.~\ref{fig:mass_spectrumProbe}, in which we compare the correct results for the spectrum of scalars, normalised to the lightest spin-2 state, with the results of the probe approximation. As expected, for most of the scalar fluctuations, and in the largest part of the parameter space of the theory, the probe approximation provides reasonably good estimates of the spectrum. But there are two important exceptions. First, we notice that the second-lightest state of the theory is never reproduced well. Even for the largest possible values of $\vr_0$, for which all other scalar states have masses that are close to the approximation, this special state has mass that is half of the estimate coming from the probe approximation. This feature persists over the whole parameter space, and becomes dramatic for $\vr_0 =\vr_0^{\rm min} \lsim 0.785$, in which case the probe approximation yields an unphysical tachyon. The inevitable conclusion of this analysis is that the second to lightest scalar state (and not the lightest!) is a dilaton, the PNGB associated with the spontaneous breaking of scale invariance.

Second, we notice that for $\vr_0\lsim \vr_0^{\rm min}$ other pathological behaviors develop in the probe approximation, besides the appearance of a dilaton. In this extreme region of parameter space, both the lightest scalars, which are almost degenerate in mass, have mass that is not reproduced in the probe approximation. But also the heavier states show visible discrepancies, even the approximate degeneracies being reproduced only poorly. We conclude that in this region of parameter space mixing effects are important for the whole spectrum of scalars, all of which couple to the dilatation operator in the dual field theory. As discuss above, in this region we also expect the supergravity approximation to be less reliable, and hence we take the value $\vr_0\sim \vr_0^{\rm min}$, at which a tachyon appears in the probe approximation, as an estimate of the lower value of $\vr_0$ for which the soliton  solutions with non-trivial flux can be trusted to yield physically reliable information about the observables in the confining dual field theory.

\section{Summary and Outlook}
\label{Sec:outlook}

We have reported our results for an extensive stability analysis of the new class of supergravity solutions recently constructed and  reported in Ref.~\cite{Fatemiabhari:2024aua}. The backgrounds arise in the context of Romans half-maximal supergravity in six dimensions. They are characterised by the compactification of one of the external directions of the geometry on a circle. The diagonal, $U(1)$ subgroup of the $SU(2)$ gauge theory develops a non-trivial flux along this compact direction. We confirmed that this construction gives rise to a one-parameter family of new solutions, that can be labelled by the value of the end of space, $\frac{3}{4}\leq \vr_0 \leq \frac{9}{10}$, expressed in a particularly convenient choice of variable for the holographic direction of the geometry. In this variable,  the background solutions can be written in analytical, closed form. The corresponding background geometries are smooth and regular,  except for the extreme $\vr_0\rightarrow \frac{3}{4}$. We confirmed that these solutions correspond to confining field theories with vacuum free energy lower than the associated conformal field theories obtained with the same deformation parameters. The only exception is the case $\vr_0=\frac{3}{4}$, for which both (inequivalent) confining and conformal vacua have vanishing free energy. The coexistence of phases is a strong indication of the existence of a first-order phase transition.

We performed a local stability analysis, by computing the spectrum of fluctuations of the fields retained in the sub-truncation of gravity theory that we described in the main body of the paper.  The results for the fluctuations are interpreted in terms of the mass spectrum of bound states of the dual field theory. We found evidence of the existence of two, nearly degenerate scalar bound states with mass lower than the typical scale of all other states, and of the confinement scale, that we identify with $M_2$, the mass of the lightest spin-2 state. We find that the next-to-lightest state behaves as a dilaton, while the lightest state does not, by repeating the calculation of the spectrum in the probe approximation. We also find that the probe approximation yields an unphysical tachyon for $\vr_0 \lsim \vr_0^{\rm min} \sim 0.785$, which we take as an estimate of the lowest value of $\vr_0$ for which the supergravity approximation can be trusted. In the whole region of parameter space, $\vr_0^{\rm min} \lsim \vr_0 \leq 9/10$, over which our analysis yields reliable estimates of the field theory observables, we find that the two lightest scalars (one of which is identified with the dilaton) have near-degenerate masses, in the range $(0.3\div 0.6)M_2$.

As anticipated in the introduction, this top-down holographic construction exemplifies a behaviour that resembles, in many respects, those presented by the theories discussed in Refs.~\cite{Elander:2020ial,Elander:2020fmv,Elander:2021wkc} and~\cite{Elander:2022ebt,Fatemiabhari:2024lct}. The presence of a first-order phase transition appears to be related with an ${\cal O}(1)$ suppression of the mass of the dilaton in the spectrum of bound states. But this suppression is neither parametrically nor numerically a large effect. This study also presents some new, unexpected feature, in the fact that the dilaton is not the lightest state, anywhere in parameter space, as an additional scalar particle with near-degenerate but smaller mass is also present. Furthermore, evidence of mixing between these two states and of parametric suppression of the mass appears in a high-curvature region of the parameter space.

It would be interesting to resolve the singularity that appears for the one background at the end of the one-parameter family of solutions discussed in this paper. Doing so might require considering a more general ansatz for the metric, the sigma-model scalars and the fluxes, which might prove challenging. An alternative way to further develop this study in the future would be to understand whether it is possible to generalise the family of solutions to explore a region of parameter space in which the background is weakly curved and regular, yet the phase transition is weaker. It should be possible to do so by turning on small deformations of the background, with the parameter $\phi_{\alpha}\neq 0$, for intermediate values of $\vr_0$ away from singularities. This process is reminiscent of what happens when adding (heavy) fermion matter fields to Yang-Mills theories on the lattice, in which case it is expected that the first-order phase transition soften into a critical line with an end point, followed by a cross-over. Finally, the same construction exploited here, generalising the mechanism for confinement of Witten's model~\cite{Witten:1998zw} by adding a non-trivial Abelian flux~\cite{Anabalon:2021tua}, can be explored in other theories and in other dimensions. It would be useful to have a complete classification of what are the physical implications of such procedure. We leave these tasks for future investigation.


\begin{acknowledgments}

AF and MP are grateful to Carlos Nunez and James Rucinski for useful discussions, and to Daniel Elander for comments, and for noticing a mistake in one of the equations of earlier versions of the manuscript.

AF has been supported by the UKRI Grant UKRI3028.

The work of MP has been supported in part by the STFC  Consolidated Grants 
No.  ST/T000813/1 and No. ST/X000648.
MP received funding from the European Research Council (ERC) under the European 
Union’s Horizon 2020 research and innovation program under Grant Agreement No.~813942.

\vspace{1.0cm}
{\bf Research Data Access Statement}---The data generated for this manuscript can be downloaded from  Ref.~\cite{data_release}. 
\vspace{1.0cm}

{\bf Open Access Statement}---For the purpose of open access, the authors have applied a Creative Commons 
Attribution (CC BY) licence  to any Author Accepted Manuscript version arising.

\end{acknowledgments}


\appendix

\section{Sigma-model and gauge invariance}
\label{Sec:sigma}

In this Appendix, we repeat the equations governing a general sigma-model of $n$ scalars coupled to gravity, including in the action terms with up to two space-time derivatives, borrowing material from Refs.~\cite{Elander:2010wd,Elander:2010wn,Bianchi:2003ug,Berg:2005pd,Berg:2006xy,Elander:2009bm,Elander:2014ola,Elander:2018aub,
Elander:2020csd,Elander:2024lir,Piai:2026rst}. In $D$ space-time dimensions,\footnote{In the body of this paper, we set $D=5$, though we also include some vector fields in addition to the scalars} the sigma-model
action, for real scalar fields $\Phi^a$, with $a=1,\,\cdots,\,n$, can be written as
\beqs
\label{eq:sigmaaction}
\mathcal S_D &=& \int \dd^D x \sqrt{-g_D} \, \left\{ \frac{\cal R}{4} - \frac{1}{2} g^{MN} G_{ab} \partial_M \Phi^a \partial_N \Phi^b - \mathcal V(\Phi^a) \right\}\nonumber \,,
\eeqs
where $G_{ab}$ is the  sigma-model  metric,  acting on the internal space, while $\mathcal V(\Phi^a)$ is a scalar potential. For gravity, we adopt conventions in which the Christoffel symbols are defined by the relations
\beqs
\Gamma^{P}{}_{MN}
&\equiv& \tfrac12\, g^{PQ}
\left(
\partial_{M} g_{NQ}
+ \partial_{N} g_{MQ}
- \partial_{Q} g_{MN}
\right)\,.
\eeqs
The tensors entering gravity formalism are defined from these, in the following way. The Riemann tensor is
\begin{equation}
\cR_{MNP}{}^Q \equiv \partial_{N}\Gamma^{Q}{}_{MP}
- \partial_{M}\Gamma^{Q}{}_{NP}
+ \Gamma^{S}{}_{MP}\Gamma^{Q}{}_{SN}
- \Gamma^{S}{}_{NP}\Gamma^{Q}{}_{SM}\,,
\end{equation}
 the Ricci tensor is obtained by contracting the indices, as
$
\cR_{MN}
= \cR_{MPN}{}^P
$, and the Ricci scalar is  $
\cR = \cR_{MN} g^{MN}$.

In analogy with gravity, one introduces the sigma-model connection, that  descends from the sigma-model metric, $ G_{ab}$,  its inverse, $G^{ab}$ and the sigma-model derivative,  $\partial_b=\frac{\partial}{\partial \Phi^b}$,  as follows:
\beqs
 {\cal G}^d_{\,\,\,\,ab}&\equiv& \frac{1}{2} G^{dc}\left(\frac{}{}\partial_a G_{cb}
+\partial_b G_{ca}-\partial_c G_{ab}\right)\,.
\eeqs
The sigma-model Riemann tensor, capturing the curvature of the internal sigma-model space, is
\beqs
 {\cal R}^a_{\,\,\,\,bcd}
&\equiv& \partial_c {\cal G}^a_{\,\,\,\,bd}-\partial_d {\cal G}^a_{\,\,\,\,bc}
+ {\cal G}^e_{\,\,\,\,bd} {\cal G}^a_{\,\,\,\,ce}- {\cal G}^e_{\,\,\,\,bc} {\cal G}^a_{\,\,\,\,de}\,.
\eeqs
The sigma-model covariant derivative is then
\beqs
D_b X^d_{\,\,\,\,a}&\equiv& \partial_b X^d_{\,\,\,\,a}+{\cal G}^d_{\,\,\,\,cb}X^c_{\,\,\,\,a}
- {\cal G}^c_{\,\,\,\,ab}X^d_{\,\,\,\,c}\,.
\eeqs

We write the domain-wall ansatz for the metric in the following form:
\beqs
\di s_D^2 &=& e^{2A} \di x_{1,D-2}^2 + \di r^2\,,
\eeqs
where the background is characterised by the function $A=A(r)$, dependent only on the holographic direction in the geometry. The short-hand $\di x_{1,D-2}^2$ stands for the flat Minkowski metric in $D-1$ dimensions, with Lorentzian signature mostly $+$. Likewise, the background scalar fields  have non-trivial profiles that depend only on the holographic direction, as $\Phi^a =  \Phi^a(r)$. The background equations take the following form:
\beqs
\label{eq:EOM1}
\partial_r^2\Phi^a\,+\,(D-1)\partial_r {A}\partial_r\Phi^a\,+\, {\cal G}^a_{\,\,\,\,bc}\partial_r\Phi^b\partial_r\Phi^c\,-\,\mathcal V^a
&=&0\,,\\
\label{eq:EOM2}
(D-1)(\partial_r { A})^2\,+\,\partial_r^2 { A}\,+\,\frac{4}{D-2} \mathcal V &=&
0\,,\\
\label{eq:EOM3}
(D-1)(D-2)(\partial_r { A})^2\,-\,2 G_{ab}\partial_r\Phi^a\partial_r\Phi^b\,+\,4 \mathcal V&=&0\,,
\eeqs
where we use the shorthand notation $\mathcal V^a\equiv  G^{ab}\partial_b \mathcal V$, and  
$\partial_b \mathcal V\equiv \frac{\partial \mathcal V}{\partial \Phi^b}$.

\subsection{Gauge invariant formalism for the fluctuations}
\label{Sec:gaugeinvariantformalism}

The equations that govern the gauge-invariant treatment of the fluctuations are taken from Refs.~\cite{Bianchi:2003ug,Berg:2005pd,Berg:2006xy,Elander:2009bm,Elander:2010wd,Elander:2010wn,Elander:2014ola,Elander:2018aub,Elander:2020csd}.  We rewrite the general scalar, $\Phi^a$, by splitting it in background and (small) fluctuations,  following Refs.~\cite{Bianchi:2003ug,Berg:2006xy,Berg:2005pd,Elander:2009bm,Elander:2010wd}, to read:
\beqs
	\Phi^a(x^\mu,r) &=&  \Phi^a(r) + \varphi^a(x^\mu,r) \,,
\eeqs
where $\varphi^a(x^\mu,r)$ are small fluctuations around the background 
solutions, $\Phi^a(r)$.
After foliating the holographic direction as in the Arnowitt-Deser-Misner  (ADM) formalism~\cite{Arnowitt:1959ah,Arnowitt:1962hi}, the metric reads as follows
\beqs
	\dd s_D^2 &=& \left( (1 + \nu)^2 + \nu_\sigma \nu^\sigma \right) \dd r^2 + 2 \nu_\mu \dd x^\mu \dd r  + e^{2 {A}(r)} \left( \eta_{\mu\nu} + h_{\mu\nu} \right) \dd x^\mu \dd x^\nu \,,
\eeqs
where we decompose the fluctuations of the metric in $D-1$ dimensions in terms of fields $\nu(x^\mu,r)$, $\nu^\mu(x^\mu,r)$,  $\mathfrak e^\mu{}_\nu(x^\mu,r)$, $\epsilon^\mu(x^\mu,r)$, $H(x^\mu,r)$, and $h(x^\mu,r)$, defined so that 
\beqs
	h^\mu{}_\nu &\equiv& \mathfrak e^\mu{}_\nu + i q^\mu \epsilon_\nu + i  \epsilon^\mu q_\nu + \frac{q^\mu q_\nu}{q^2} H + \frac{1}{D-2} \delta^\mu{}_\nu h\,.
\eeqs
(In the body of the paper, we specify that $D=5$.) In these expressions, the gauge-invariant, vector field, $\epsilon^\mu(x^\mu,r)$,  and the traceless  tensor, $\mathfrak e^\mu{}_\nu(x^\mu,r)$,
 are transverse. Following Ref.~\cite{Berg:2005pd,Berg:2006xy}, the other gauge-invariant combinations  of the fluctuations are given by
\beqs
\label{eq:scalarfluc}
\mathfrak a^a &\equiv& \varphi^a - \frac{\partial_r  \Phi^a}{2(D-2)\partial_r { A}} h \,, \\
\mathfrak b &\equiv& \nu - \partial_r \left( \frac{h}{2(D-2)\partial_r {A}} \right) \,, \\
\mathfrak c &\equiv& e^{-2{ A}} \partial_\mu \nu^\mu - \frac{e^{-2{ A}} q^2 h}{2(D-2) \partial_r A} 
	- \frac{1}{2} \partial_r H \,, \\
\mathfrak d^\mu &\equiv& e^{-2{A}} P^\mu{}_\nu \nu^\nu - \partial_r \epsilon^\mu \,.
\eeqs
The algebraic equations obeyed by $\mathfrak{b}$, $\mathfrak{c}$, and $\mathfrak{d}$ allow to decouple them and do not lead to a spectrum of mass eigenstates.

In our numerical calculations, we assume the holographic direction to be bounded, as in $r_1<r<r_2$. 
 It is understood that the limits $r_1\rightarrow r_0$ and $r_2\rightarrow +\infty$ are taken at the end of the calculation, to remove the spurious dependence on the regulators. The tensor fluctuations, $\mathfrak e^\mu{}_\nu$, obey  comparatively simple differential equations:
\beq
\label{eq:teom}
	\left[ \partial_r^2 + (D-1) \partial_r {A} \partial_r - e^{-2{ A}(r)} q^2 \right] \mathfrak e^\mu_{\,\,\,\nu} = 0 \,,
\eeq
and their (Neumann) boundary conditions are given by 
\beq
\label{eq:tbc}
\left.\frac{}{}	\partial_r \mathfrak e ^\mu_{\,\,\,\nu} \right|_{r=r_i}= 0 \,.
\eeq

The equations of motion for the scalar fluctuations, $\mathfrak a^a$, are the following:
\beqs
\label{eq:seom}
	0 &=& \Big[ {\cal D}_r^2 + (D-1) \partial_{r}{A} {\cal D}_r - e^{-2{A}} q^2 \Big] \mathfrak{a}^a \,\,\\ \nonumber
	&& - \Big[  {\mathcal V}^{\,\,a}{}_{\,|c} - \mathcal{R}^a{}_{bcd} \partial_{r}\Phi^b \partial_{r}\Phi^d + 
	\frac{4 (\partial_{r}\Phi^a  {\mathcal V}^{\,b} +  {\mathcal V}^{\,a} 
	\partial_{r}\Phi^b) G_{bc}}{(D-2) \partial_{r} {A}} + 
	\frac{16  {\mathcal V} \partial_{r}\Phi^a \partial_{r}\Phi^b G_{bc}}{(D-2)^2 (\partial_{r}{A})^2} \Big] \mathfrak{a}^c\,,
\eeqs
where $q^2=\eta_{\mu\nu}q^{\mu}q^{\nu}=-M^2$, the background covariant derivative is  
$\mathcal D_r \mathfrak a^a \equiv \partial_r \mathfrak a^a +
 \mathcal G^a_{\ bc} \partial_r  \Phi^b \mathfrak a^c$,
and ${\mathcal V}^a{}_{|b} \equiv \frac{\partial {\mathcal V}^a}{\partial \Phi^b} + \mathcal G^a_{\ bc} {\mathcal V}^c$.
The boundary conditions are~\cite{Elander:2010wd}
\beqs
\label{eq:sbc}
 \frac{2  e^{2A}\partial_{r} \Phi^a }{(D-2)q^2 \partial_{r}{ A}}
	\left[ \partial_{r} \Phi^b{\cal D}_r -\frac{4  {\cal V} \partial_{r} \Phi^b}{(D-2) 
	\partial_r { A}} - {\cal V}^b \right] \mathfrak a_b \Big|_{r_i} &=& \mathfrak a^a\Big|_{r_i}  \, .
\eeqs

\subsubsection{Probe approximation}
\label{Sec:probeEquations}

The probe approximation in use in the body of the paper is defined in Ref.~\cite{Elander:2020csd}, and relies on the assumption that the general gauge-invariant scalar fluctuation can be expanded in a power series of the small quantities, $\frac{\partial_r\Phi^a}{\partial_r A} \ll 1$, that control the mixing between the fluctuations of the sigma-model scalars, $\varphi^a$, and the trace of the metric, $h$, in Eq.~(\ref{eq:scalarfluc}). By truncating at the leading order, the equations written in the probe approximation simplify, to read
\beqs
\label{Eq:probeeomgeneral}
0&=&\left[\frac{}{}{\cal D}_r^2 + 4\partial_r A\,{\cal D}_r -e^{-2A} q^2\right]\mathfrak{p}^a-\left[\frac{}{}V^a_{\,\,\,\,|c}-{\cal R}^a_{\,\,\,\,bcd}\partial_r\Phi^b\partial_r \Phi^d
\right]\mathfrak{p}^c\,,
\eeqs
where we denoted the relevant variable as $\mathfrak{p}^a$, rather than $\mathfrak{a}^a$, to avoid confusion. The boundary conditions reduce to Dirichlet: $\left.\frac{}{}\mathfrak{p}^a\right|_{r_i}=0$. In the numerical calculations, we find it convenient to change the variable describing the holographic direction from $r$ to $\vr$,  as we have the background fields written as simple function of the coordinate $\vr$.

\subsection{Equations of motion for fluctuations}\label{Sec:eom}

In this Appendix, we write explicitly the equations of motion for the fluctuations of the scalars, both in gauge-invariant form, and in the probe approximation.  For simplicity, we write the equation in the coordinate $r$, although in the numerical study we performed we used the change of variable to $\vr$.  We also find it convenient to denote derivatives in respect to $r$ with primed functions, so that $A^{\prime}(r)\equiv \partial_{r}A(r)$, $A^{\prime\prime}(r)\equiv \partial_{r}^2A(r)$, and similarly for all other functions. We performed the replacement $q^2=-M^2$.

\subsubsection{Gauge-invariant scalar fluctuations} 
The coupled, second-order, linear differential equations obeyed by  the three gauge-invariant combinations of scalar fluctuations, $\mathfrak{a}^{\phi}$, $\mathfrak{a}^{\chi}$, and  $\mathfrak{a}^{A_6}$,  are the following:
\begin{align}
    &\frac{1}{162 A'(r)^2}e^{-2 (A(r)+4 (\chi (r)+\phi (r)))} \left(e^{2 A(r)} \left(81 A'(r)^2 e^{2 \chi (r)+6 \phi (r)} \left(2 \mathfrak{a}^{\phi \prime\prime}(r) e^{6 \chi (r)+2 \phi (r)}+2 \mathfrak{a}^{A_6 \prime}(r) A_6^{(3) \prime}(r)+\right.\right.\right.\nonumber\\
    &\left.\left.\left.
    \mathfrak{a}^{A_6}(r) \left(A_6^{(3) \prime\prime}(r)-2 A_6^{(3) \prime}(r) \left(3 \chi '(r)+\phi '(r)\right)\right)\right)+324 A'(r)^3 e^{2 \chi (r)+6 \phi (r)} \left(2 \mathfrak{a}^{\phi \prime}(r) e^{6 \chi (r)+2 \phi (r)}+\mathfrak{a}^{A_6}(r) A_6^{(3) \prime}(r)\right)\right.\right.\nonumber\\
    &\left.\left.-6 \mathfrak{a}^{\chi}(r) e^{2 (\chi (r)+\phi (r))} \left(9 A'(r)^2 \left(9 e^{4 \phi (r)} A_6^{(3) \prime}(r)^2+2 e^{4 \chi (r)} \left(-4 e^{4 \phi (r)}+3 e^{8 \phi (r)}+1\right)\right)\right.\right.\right.\nonumber\\
    &\left.\left.\left.+8 e^{4 \chi (r)} A'(r) \left(\left(12 e^{4 \phi (r)}+9 e^{8 \phi (r)}-1\right) \phi '(r)-9 \left(-4 e^{4 \phi (r)}+3 e^{8 \phi (r)}+1\right) \chi '(r)\right)\right.\right.\right.\nonumber\\
    &\left.\left.\left.-32 e^{4 \chi (r)} \left(12 e^{4 \phi (r)}+9 e^{8 \phi (r)}-1\right) \chi '(r) \phi '(r)\right)+72 \mathfrak{a}^{A_6}(r) \left(-4 e^{4 \phi (r)}+3 e^{8 \phi (r)}+1\right) A'(r) A_6^{(3) \prime}(r)\right.\right.\nonumber\\
    &\left.\left.+32 \mathfrak{a}^{A_6}(r) \left(12 e^{4 \phi (r)}+9 e^{8 \phi (r)}-1\right) A_6^{(3) \prime}(r) \phi '(r)\right)\right.\nonumber\\
    &\left.+2 \mathfrak{a}^{\phi}(r) e^{2 (\chi (r)+\phi (r))} \left(27 A'(r)^2 \left(e^{4 \chi (r)} \left(8 e^{2 A(r)+4 \phi (r)}+6 e^{2 A(r)+8 \phi (r)}-6 e^{2 A(r)}+3 M^2 e^{2 \chi (r)+6 \phi (r)}\right)\right.\right.\right.\nonumber\\
    &\left.\left.\left.-3 e^{2 A(r)+4 \phi (r)} A_6^{(3) \prime}(r)^2\right)+144 \left(-4 e^{4 \phi (r)}+3 e^{8 \phi (r)}+1\right) e^{2 A(r)+4 \chi (r)} A'(r) \phi '(r)\right.\right.\nonumber\\
    &\left.\left.+32 \left(12 e^{4 \phi (r)}+9 e^{8 \phi (r)}-1\right) e^{2 A(r)+4 \chi (r)} \phi '(r)^2\right)\right)=0\,,
    \end{align}
\begin{align}
   &\frac{1}{162 A'(r)^2} e^{-2 (A(r)+4 (\chi (r)+\phi (r)))} \left(-2 \mathfrak{a}^{\phi}(r) e^{2 (A(r)+\chi (r)+\phi (r))} \left(9 A'(r)^2 \left(9 e^{4 \phi (r)} A_6^{(3) \prime}(r)^2\right.\right.\right.\nonumber\\
    &\left.\left.\left.+2 e^{4 \chi (r)} \left(-4 e^{4 \phi (r)}+3 e^{8 \phi (r)}+1\right)\right)+8 e^{4 \chi (r)} A'(r) \left(\left(12 e^{4 \phi (r)}+9 e^{8 \phi (r)}-1\right) \phi '(r)-\right.\right.\right.\nonumber\\
    &\left.\left.\left.9 \left(-4 e^{4 \phi (r)}+3 e^{8 \phi (r)}+1\right) \chi '(r)\right)-32 e^{4 \chi (r)} \left(12 e^{4 \phi (r)}+9 e^{8 \phi (r)}-1\right) \chi '(r) \phi '(r)\right)\right.\nonumber\\
    &\left.+e^{2 A(r)} \left(81 A'(r)^2 e^{2 \chi (r)+6 \phi (r)} \left(2 \mathfrak{a}^{\chi \prime\prime}(r) e^{6 \chi (r)+2 \phi (r)}+2 \mathfrak{a}^{A_6 \prime}(r) A_6^{(3) \prime}(r)+\right.\right.\right.\nonumber\\
    &\left.\left.\left.\mathfrak{a}^{A_6}(r) \left(A_6^{(3) \prime\prime}(r)-2 A_6^{(3) \prime}(r) \left(3 \chi '(r)+\phi '(r)\right)\right)\right)+324 A'(r)^3 e^{2 \chi (r)+6 \phi (r)} \left(2 \mathfrak{a}^{\chi \prime}(r) e^{6 \chi (r)+2 \phi (r)}+\mathfrak{a}^{A_6}(r) A_6^{(3) \prime}(r)\right)\right.\right.\nonumber\\
    &\left.\left.-8 \mathfrak{a}^{A_6}(r) \left(12 e^{4 \phi (r)}+9 e^{8 \phi (r)}-1\right) A'(r) A_6^{(3) \prime}(r)+32 \mathfrak{a}^{A_6}(r) \left(12 e^{4 \phi (r)}+9 e^{8 \phi (r)}-1\right) A_6^{(3) \prime}(r) \chi '(r)\right)\right.\nonumber\\
    &\left.+6 \mathfrak{a}^{\chi}(r) e^{2 (\chi (r)+\phi (r))} \left(A'(r)^2 \left(e^{4 \chi (r)} \left(24 e^{2 A(r)+4 \phi (r)}+18 e^{2 A(r)+8 \phi (r)}-2 e^{2 A(r)}+27 M^2 e^{2 \chi (r)+6 \phi (r)}\right)\right.\right.\right.\nonumber\\
    &\left.\left.\left.-81 e^{2 A(r)+4 \phi (r)} A_6^{(3) \prime}(r)^2\right)-16 \left(12 e^{4 \phi (r)}+9 e^{8 \phi (r)}-1\right) e^{2 A(r)+4 \chi (r)} A'(r) \chi '(r)\right.\right.\nonumber\\
    &\left.\left.+32 \left(12 e^{4 \phi (r)}+9 e^{8 \phi (r)}-1\right) e^{2 A(r)+4 \chi (r)} \chi '(r)^2\right)\right)=0\,,
   \end{align}
\begin{align}
  & 4 A'(r) \left(-\mathfrak{a}^{\phi}(r) A_6^{(3) \prime}(r)-3 \mathfrak{a}^{\chi}(r) A_6^{(3) \prime}(r)\right.\nonumber\\
    &\left.+\mathfrak{a}^{A_6 \prime}(r)-3 \mathfrak{a}^{A_6}(r) \chi '(r)-\mathfrak{a}^{A_6}(r) \phi '(r)\right)+\frac{1}{81 A'(r)^2}\left(16 \left(12 e^{4 \phi (r)}+9 e^{8 \phi (r)}-1\right) A_6^{(3) \prime}(r) e^{-8 (\chi (r)+\phi (r))} \right.\nonumber\\
    &\left.\left(2 e^{6 \chi (r)+2 \phi (r)} \left(\mathfrak{a}^{\phi}(r) \phi '(r)+3 \mathfrak{a}^{\chi}(r) \chi '(r)\right)+\mathfrak{a}^{A_6}(r) A_6^{(3) \prime}(r)\right)\right)\nonumber\\
    &+\frac{8 A_6^{(3) \prime}(r) e^{-2 (\chi (r)+3 \phi (r))} \left(3 \mathfrak{a}^{\phi}(r) \left(-4 e^{4 \phi (r)}+3 e^{8 \phi (r)}+1\right)-\mathfrak{a}^{\chi}(r) \left(12 e^{4 \phi (r)}+9 e^{8 \phi (r)}-1\right)\right)}{27 A'(r)}\nonumber\\
    &+M^2 e^{-2 A(r)} \mathfrak{a}^{A_6}(r)-2 \mathfrak{a}^{\phi \prime}(r) A_6^{(3) \prime}(r)-\mathfrak{a}^{\phi}(r) A_6^{(3) \prime\prime}(r)+6 \mathfrak{a}^{\phi}(r) A_6^{(3) \prime}(r) \chi '(r)+2 \mathfrak{a}^{\phi}(r) A_6^{(3) \prime}(r) \phi '(r)\nonumber\\
    &-6 \mathfrak{a}^{\chi \prime}(r) A_6^{(3) \prime}(r)-3 \mathfrak{a}^{\chi}(r) A_6^{(3) \prime\prime}(r)+18 \mathfrak{a}^{\chi}(r) A_6^{(3) \prime}(r) \chi '(r)+6 \mathfrak{a}^{\chi}(r) A_6^{(3) \prime}(r) \phi '(r)+\mathfrak{a}^{A_6 \prime\prime}(r)\nonumber\\
    &-6 \mathfrak{a}^{A_6 \prime}(r) \chi '(r)-2 \mathfrak{a}^{A_6 \prime}(r) \phi '(r)-2 \mathfrak{a}^{A_6}(r) A_6^{(3) \prime}(r)^2 e^{-2 (3 \chi (r)+\phi (r))}-3 \mathfrak{a}^{A_6}(r) \chi ''(r)\nonumber\\
    &-\mathfrak{a}^{A_6}(r) \phi ''(r)+\frac{4}{9} \mathfrak{a}^{A_6}(r) \left(6 e^{4 \phi (r)}-1\right) e^{-2 (\chi (r)+3 \phi (r))}=0\,.
\end{align}

The gauge-invariant scalar fluctuations obey the following boundary conditions, which we impose at the cutoffs $r_1$ (in the IR) and $r_2$ (in the UV), along the holographic direction:
\begin{align}
   & \frac{2}{27} e^{2 A(r)} \phi '(r) \left(\frac{4 \left(12 e^{4 \phi (r)}+9 e^{8 \phi (r)}-1\right) e^{-8 (\chi (r)+\phi (r))} \left(2 e^{6 \chi (r)+2 \phi (r)} \left( \mathfrak{a}^{\phi }(r) \phi '(r)+3  \mathfrak{a}^{\chi}(r) \chi '(r)\right)+ \mathfrak{a}^{A_6}(r) A_6^{(3) \prime}(r)\right)}{A'(r)} \right.\nonumber\\
    &\left.+54 \phi '(r) \left( \mathfrak{a}^{\phi \prime}(r)+\frac{1}{2}  \mathfrak{a}^{A_6}(r) A_6^{(3) \prime}(r) e^{-2 (3 \chi (r)+\phi (r))}\right) \right.\nonumber\\
    &\left.-27 A_6^{(3) \prime}(r) e^{-2 (3 \chi (r)+\phi (r))} \left( \mathfrak{a}^{\phi }(r) A_6^{(3) \prime}(r)+3  \mathfrak{a}^{\chi}(r) A_6^{(3) \prime}(r)- \mathfrak{a}^{A_6 \prime}(r)+3  \mathfrak{a}^{A_6}(r) \chi '(r)+ \mathfrak{a}^{A_6}(r) \phi '(r)\right) \right.\nonumber\\
    &\left.+18  \mathfrak{a}^{\phi }(r) \left(-4 e^{4 \phi (r)}+3 e^{8 \phi (r)}+1\right) e^{-2 (\chi (r)+3 \phi (r))}+162 \chi '(r) \left( \mathfrak{a}^{\chi \prime}(r)+\frac{1}{2}  \mathfrak{a}^{A_6}(r) A_6^{(3) \prime}(r) e^{-2 (3 \chi (r)+\phi (r))}\right) \right.\nonumber\\
    &\left.\frac{}{}\left.-6  \mathfrak{a}^{\chi}(r) \left(12 e^{4 \phi (r)}+9 e^{8 \phi (r)}-1\right) e^{-2 (\chi (r)+3 \phi (r))}\right)+3 M^2  \mathfrak{a}^{\phi }(r) A'(r)
    \right|_{r=r_i}=0\,,
\end{align}
\begin{align}
    &\frac{2}{27} e^{2 A(r)} \chi '(r) \left(\frac{4 \left(12 e^{4 \phi (r)}+9 e^{8 \phi (r)}-1\right) e^{-8 (\chi (r)+\phi (r))} \left(2 e^{6 \chi (r)+2 \phi (r)} \left( \mathfrak{a}^{\phi }(r) \phi '(r)+3  \mathfrak{a}^{\chi}(r) \chi '(r)\right)+ \mathfrak{a}^{A_6}(r) A_6^{(3) \prime}(r)\right)}{A'(r)} \right.\nonumber\\
    &\left.+54 \phi '(r) \left( \mathfrak{a}^{\phi \prime}(r)+\frac{1}{2}  \mathfrak{a}^{A_6}(r) A_6^{(3) \prime}(r) e^{-2 (3 \chi (r)+\phi (r))}\right) \right.\nonumber\\
    &\left.-27 A_6^{(3) \prime}(r) e^{-2 (3 \chi (r)+\phi (r))} \left( \mathfrak{a}^{\phi }(r) A_6^{(3) \prime}(r)+3  \mathfrak{a}^{\chi}(r) A_6^{(3) \prime}(r)- \mathfrak{a}^{A_6 \prime}(r)+3  \mathfrak{a}^{A_6}(r) \chi '(r)+ \mathfrak{a}^{A_6}(r) \phi '(r)\right) \right.\nonumber\\
    &\left.+18  \mathfrak{a}^{\phi }(r) \left(-4 e^{4 \phi (r)}+3 e^{8 \phi (r)}+1\right) e^{-2 (\chi (r)+3 \phi (r))}+162 \chi '(r) \left( \mathfrak{a}^{\chi \prime}(r)+\frac{1}{2}  \mathfrak{a}^{A_6}(r) A_6^{(3) \prime}(r) e^{-2 (3 \chi (r)+\phi (r))}\right) \right.\nonumber\\
    &\left.\frac{}{}
    \left.-6  \mathfrak{a}^{\chi}(r) \left(12 e^{4 \phi (r)}+9 e^{8 \phi (r)}-1\right) e^{-2 (\chi (r)+3 \phi (r))}\right)+3 M^2  \mathfrak{a}^{\chi}(r) A'(r)    \right|_{r=r_i}=0\,,
\end{align}
\begin{align}
    &2 A_6^{(3) \prime}(r) e^{2 (A(r)+\chi (r)+\phi (r))} \left(27 e^{4 \phi (r)} A'(r) \left(2 e^{6 \chi (r)+2 \phi (r)} \left( \mathfrak{a}^{\phi \prime}(r) \phi '(r)+3  \mathfrak{a}^{\chi \prime}(r) \chi '(r)\right)+ \mathfrak{a}^{A_6 \prime}(r) A_6^{(3) \prime}(r)\right)\right.\nonumber\\
    &\left.+9  \mathfrak{a}^{\phi }(r) A'(r) \left(2 e^{4 \chi (r)} \left(-4 e^{4 \phi (r)}+3 e^{8 \phi (r)}+1\right)-3 e^{4 \phi (r)} A_6^{(3) \prime}(r)^2\right)\right.\nonumber\\
    &\left.+3  \mathfrak{a}^{\chi}(r) \left(8 e^{4 \chi (r)} \left(12 e^{4 \phi (r)}+9 e^{8 \phi (r)}-1\right) \chi '(r)-A'(r) \left(27 e^{4 \phi (r)} A_6^{(3) \prime}(r)^2+2 e^{4 \chi (r)} \left(12 e^{4 \phi (r)}+9 e^{8 \phi (r)}-1\right)\right)\right)\right.\nonumber\\
    &\left.+8  \mathfrak{a}^{\phi }(r) e^{4 \chi (r)} \left(12 e^{4 \phi (r)}+9 e^{8 \phi (r)}-1\right) \phi '(r)\right)\nonumber\\
    &+\left.\frac{}{} \mathfrak{a}^{A_6}(r) \left(81 M^2 A'(r)^2 e^{8 (\chi (r)+\phi (r))}+8 e^{2 A(r)} \left(12 e^{4 \phi (r)}+9 e^{8 \phi (r)}-1\right) A_6^{(3) \prime}(r)^2\right)    \right|_{r=r_i}=0\,.
\end{align}

\subsubsection{Probe approximation}
The equations of motion for the three scalar fluctuations, treated in the probe approximation, are the following:
\begin{align}
    &e^{2 A(r)} \left(3 e^{4 \phi (r)} \left(2 \left(4 A'(r)  \mathfrak{p}^{\phi \prime }(r) e^{6 \chi (r)+2 \phi (r)}+A_6^{(3) \prime }(r)  \mathfrak{p}^{A_6\prime}(r)+ \mathfrak{p}^{\phi \prime \prime}(r) e^{6 \chi (r)+2 \phi (r)}\right)\right.\right.\nonumber\\
    &\left.\left.+ \mathfrak{p}^{A_6}(r) \left(4 A'(r) A_6^{(3) \prime }(r)+A_6^{(3) \prime \prime}(r)-2 A_6^{(3) \prime }(r) \left(3 \chi '(r)+\phi '(r)\right)\right)\right)\right.\nonumber\\
    &\left.-2  \mathfrak{p}^{\chi }(r) \left(9 e^{4 \phi (r)} A_6^{(3) \prime }(r)^2+2 e^{4 \chi (r)} \left(-4 e^{4 \phi (r)}+3 e^{8 \phi (r)}+1\right)\right)\right)\nonumber\\
    &+2  \mathfrak{p}^{\phi }(r) \left(e^{4 \chi (r)} \left(8 e^{2 A(r)+4 \phi (r)}+6 e^{2 A(r)+8 \phi (r)}-6 e^{2 A(r)}+3 M^2 e^{2 \chi (r)+6 \phi (r)}\right)-3 e^{2 A(r)+4 \phi (r)} A_6^{(3) \prime }(r)^2\right)=0\,,
\end{align}
\begin{align}
    &27 e^{2 A(r)+4 \phi (r)} \left(2 \left(4 A'(r)  \mathfrak{p}^{\chi \prime }(r) e^{6 \chi (r)+2 \phi (r)}+A_6^{(3) \prime }(r)  \mathfrak{p}^{A_6\prime}(r)+ \mathfrak{p}^{\chi \prime \prime}(r) e^{6 \chi (r)+2 \phi (r)}\right)\right.\nonumber\\
    &\left.+ \mathfrak{p}^{A_6}(r) \left(4 A'(r) A_6^{(3) \prime }(r)+A_6^{(3) \prime \prime}(r)-2 A_6^{(3) \prime }(r) \left(3 \chi '(r)+\phi '(r)\right)\right)\right)\nonumber\\
    &+2  \mathfrak{p}^{\chi }(r) \left(e^{4 \chi (r)} \left(24 e^{2 A(r)+4 \phi (r)}+18 e^{2 A(r)+8 \phi (r)}-2 e^{2 A(r)}+27 M^2 e^{2 \chi (r)+6 \phi (r)}\right)-81 e^{2 A(r)+4 \phi (r)} A_6^{(3) \prime }(r)^2\right)\nonumber\\
    &-6 e^{2 A(r)}  \mathfrak{p}^{\phi }(r) \left(9 e^{4 \phi (r)} A_6^{(3) \prime }(r)^2+2 e^{4 \chi (r)} \left(-4 e^{4 \phi (r)}+3 e^{8 \phi (r)}+1\right)\right)=0\,,
\end{align}
\begin{align}
    & \mathfrak{p}^{A_6}(r) \left(-4 A'(r) \left(3 \chi '(r)+\phi '(r)\right)+M^2 e^{-2 A(r)}-2 A_6^{(3) \prime }(r)^2 e^{-2 (3 \chi (r)+\phi (r))}-3 \chi ''(r)-\phi ''(r)\right.\nonumber\\
    &\left.+\frac{8}{3} e^{-2 (\chi (r)+\phi (r))}-\frac{4}{9} e^{-2 (\chi (r)+3 \phi (r))}\right)-12  \mathfrak{p}^{\chi }(r) A'(r) A_6^{(3) \prime }(r)\nonumber\\
    &- \mathfrak{p}^{\phi }(r) \left(4 A'(r) A_6^{(3) \prime }(r)+A_6^{(3) \prime \prime}(r)-2 A_6^{(3) \prime }(r) \left(3 \chi '(r)+\phi '(r)\right)\right)+4 A'(r)  \mathfrak{p}^{A_6\prime}(r)-3  \mathfrak{p}^{\chi }(r) A_6^{(3) \prime \prime}(r)\nonumber\\
    &-2 A_6^{(3) \prime }(r)  \mathfrak{p}^{\phi \prime }(r)-6 A_6^{(3) \prime }(r)  \mathfrak{p}^{\chi \prime }(r)+18  \mathfrak{p}^{\chi }(r) A_6^{(3) \prime }(r) \chi '(r)\nonumber\\
    &+6  \mathfrak{p}^{\chi }(r) A_6^{(3) \prime }(r) \phi '(r)+ \mathfrak{p}^{A_6\prime\prime}(r)-6  \mathfrak{p}^{A_6\prime}(r) \chi '(r)-2  \mathfrak{p}^{A_6\prime}(r) \phi '(r)=0\,,
\end{align}
where $\left\{\mathfrak{p}^{\phi},\,\mathfrak{p}^{\chi},\,\mathfrak{p}^{A_6}\right\}$ are the approximations of the gauge-invariant combinations $\left\{\mathfrak{a}^{\phi},\,\mathfrak{a}^{\chi},\,\mathfrak{a}^{A_6}\right\}$, associated to the fields $\left\{\phi,\,\chi,\,A_6^{(3)}\right\}$. All three scalar fluctuations obey Dirichlet boundary conditions:  $\left.\frac{}{}\mathfrak{p}^{\phi}\right|_{r=r_i}=\left.\frac{}{}\mathfrak{p}^{\chi}\right|_{r=r_i}=   \left.\frac{}{}\mathfrak{p}^{A_6}\right|_{r=r_i}=0$.

\section{Asymptotic expansion of the fluctuations}
\label{Sec:UVfluc}

As explained in the main body of the paper, in order to improve the convergence of our numerical algorithms for computing the spectra of spin-0 and spin-1 fluctuations, we use UV and IR expansions of the fluctuations, impose the boundary conditions on such expansions to fix the relation between the coefficients of the dominant and subdominant contribution to the expansion, and then use the resulting expansions to set the boundary conditions for the evolution of the differential equations of the gauge-invariant fluctuations. We report here these expansions, truncated at the order we used in the numerical study. 

\subsection{IR expansions}
We find it convenient to write the expansions in the IR in the variable $\vr$.
The leading terms of the IR  expansion, in powers of the small quantity $\vr-\vr_0$, of the scalar fluctuations are given by the following expressions:
\begin{equation}
    \begin{split}
    \mathfrak{a}^{\phi }_{IR}(\vr)=&\mathfrak{a}^{\phi }_{IR, 0} +\mathfrak{a}^{\phi }_{IR, L} \log(\vr-\vr_0)+ \\
    &\frac{(\vr-\vr_0)}{6 \vr_0^{5/2} (4 \vr_0-3)}\Bigg[\mathfrak{a}^{\phi }_{IR, 0} \vr_0^{3/2} \left(-2 M^2+88 \vr_0^2-174 \vr_0+81\right)+2 \mathfrak{a}^{\phi }_{IR, L} \vr_0^{3/2} \left(2 M^2-128 \vr_0^2+204 \vr_0-81\right)\\
    &-81 \mathfrak{a}^{\chi }_{IR, 0} \vr_0^{3/2}+198 \mathfrak{a}^{\chi }_{IR, 0} \vr_0^{5/2}-120 \mathfrak{a}^{\chi }_{IR, 0} \vr_0^{7/2}+162 \mathfrak{a}^{\chi }_{IR, L} \vr_0^{3/2}-288 \mathfrak{a}^{\chi }_{IR, L} \vr_0^{5/2}+120 \mathfrak{a}^{\chi }_{IR, L} \vr_0^{7/2}\\
    &+\mathfrak{a}^{A_6 }_{IR, 0} M^2 \sqrt{-\vr_0^2 (10 \vr_0-9)}+40 \mathfrak{a}^{A_6 }_{IR, 0} \sqrt{-\vr_0^2 (10 \vr_0-9)} \vr_0^2-96 \mathfrak{a}^{A_6 }_{IR, 0} \sqrt{-\vr_0^2 (10 \vr_0-9)} \vr_0\\
    &+54 \mathfrak{a}^{A_6 }_{IR, 0} \sqrt{-\vr_0^2 (10 \vr_0-9)}+12 \mathfrak{a}^{A_6 }_{IR, 1} \sqrt{-\vr_0^2 (10 \vr_0-9)} \vr_0^2-9 \mathfrak{a}^{A_6 }_{IR, 1} \sqrt{-\vr_0^2 (10 \vr_0-9)} \vr_0 \Bigg]\\
    &+\frac{\frac{\mathfrak{a}^{\phi }_{IR, L} \left(2 M^2-128 \vr_0^2+210 \vr_0-81\right)}{\vr_0}+\mathfrak{a}^{\chi }_{IR, L} \left(\frac{81}{\vr_0}-90\right)+\frac{\mathfrak{a}^{A_6 }_{IR, 0} M^2 \sqrt{-\vr_0^2 (10 \vr_0-9)}}{\vr_0^{5/2}}}{18-24 \vr_0}\log(\vr-\vr_0)(\vr-\vr_0)+\cdots\,,
    \end{split}
\end{equation}
\begin{equation}
    \begin{split}
    \ac_{IR}(\vr)=&\ac_{IR, 0} +\ac_{IR, L} \log(\vr-\vr_0)- \\
    &\frac{(\vr-\vr_0)}{18-24\vr_0}\Bigg[\mathfrak{a}^{\phi }_{IR, 0} \left(-40 \vr_0-\frac{27}{\vr_0}+66\right)+\mathfrak{a}^{\phi }_{IR, L} \left(40 \vr_0+\frac{54}{\vr_0}-96\right)+\frac{\mathfrak{a}^{\chi }_{IR, 0} \left(-2 M^2-120 \vr_0^2+54 \vr_0+27\right)}{\vr_0}+\\
    &\frac{2 \mathfrak{a}^{\chi }_{IR, L} \left(2 M^2+40 \vr_0^2+12 \vr_0-27\right)}{\vr_0}+\frac{\sqrt{-\vr_0^2 (10 \vr_0-9)} \left(\mathfrak{a}^{A_6 }_{IR, 0} \left(M^2+40 \vr_0^2-18\right)+3 \mathfrak{a}^{A_6 }_{IR, 1} \vr_0 (4 \vr_0-3)\right)}{\vr_0^{5/2}}\Bigg] +\\
    &\frac{3 \mathfrak{a}^{\phi }_{IR, L} (10 \vr_0-9) \vr_0^{3/2}+\mathfrak{a}^{\chi }_{IR, L} \vr_0^{3/2} \left(-2 M^2-54 \vr_0+27\right)-\mathfrak{a}^{A_6 }_{IR, 0} M^2 \sqrt{-\vr_0^2 (10 \vr_0-9)}}{6 \vr_0^{5/2} (4 \vr_0-3)}\log(\vr-\vr_0)(\vr-\vr_0)
    \\ &+\cdots\,,
    \end{split}
\end{equation}
\begin{equation}
    \begin{split}
        \mathfrak{a}^{A_6 }_{IR}(\vr)= &\mathfrak{a}^{A_6 }_{IR, 0}+ \mathfrak{a}^{A_6 }_{IR, 1}(\vr-\vr_0)+\log(\vr-\vr_0)(\vr-\vr_0)\Bigg[-\frac{4 \sqrt{\vr_0} \sqrt{-\vr_0^2 (10 \vr_0-9)} (\mathfrak{a}^{\phi }_{IR, L}+3 \mathfrak{a}^{\chi }_{IR, L})+\mathfrak{a}^{A_6 }_{IR, 0} M^2}{3 \vr_0 (4 \vr_0-3)}\Bigg]
        \\ & +\cdots\,,
    \end{split}
\end{equation}
where $\mathfrak{a}^{\phi }_{IR, 0} $, $\mathfrak{a}^{\phi }_{IR, L}$, $\ac_{IR, 0}$, $\ac_{IR, L}$, $\mathfrak{a}^{A_6 }_{IR, 0}$, and $\mathfrak{a}^{A_6 }_{IR, 1}$ are the six integration constants.

The leading terms of the IR expansion of the vector fluctuations are given by the following expressions:
\begin{equation}
    \begin{split}
    \tilde{A}_{IR}(\vr)&=\tilde{A}_{IR, 0} + \tilde{A}_{IR, L} \log(\vr-\vr_0)+ \frac{(\vr-\vr_0)}{9 (3-4 \vr_0)^2 \vr_0^{3/2}}\Bigg[3 \tilde{A}_{IR, 0} M^2 (3-4 \vr_0) \sqrt{\vr_0}+\\
    &-6 \tilde{A}_{IR, L} \sqrt{\vr_0} (4 \vr_0-3) \left(-M^2+30 \vr_0^2-36 \vr_0+9\right)+\\
    &+2 \sqrt{-\vr_0^2 (10 \vr_0-9)} \left(3 \vr_0 ((3-4 \vr_0) \tilde{V}_{IR, 0})+4 \vr_0(5 \vr_0-6) \tilde{V}_{IR, -1}-M^2 \tilde{V}_{IR, -1}\right)+\\
    &\log(\vr-\vr_0)\left(2 \sqrt{-\vr_0^2 (10 \vr_0-9)} \tilde{V}_{IR, -1} M^2-3 \tilde{A}_{IR, L} \sqrt{\vr_0} (4 \vr_0-3) \left(M^2+4 \vr_0 (9-10 \vr_0)\right)\right)\Bigg] \\
    &+\cdots\,,
    \end{split}
\end{equation}
\begin{equation}
    \begin{split}
    \tilde{V}_{IR}(\vr)&=\frac{\tilde{V}_{
    IR, -1}}{\vr-\vr_0}+\tilde{V}_{IR, 0}+{\log(\vr-\vr_0)}\left(\frac{2 \tilde{A}_{IR, L} \sqrt{-\vr_0^2 (10 \vr_0-9)}}{\vr_0^{3/2}}-\frac{\tilde{V}_{IR, -1}M^2}{3 \vr_0 (4 \vr_0-3)}\right)+ \\
    &\frac{(\vr-\vr_0)}{36 (3-4 \vr_0)^2 \vr_0^{5/2}}\Bigg[6 M^2 \left(2 \sqrt{\vr_0} \left(10 \vr_0^2-12 \vr_0+3\right) \tilde{V}_{IR, -1}\right.+\\
    &\left.-(4 \vr_0-3) \left(2 \sqrt{-\vr_0^2 (10 \vr_0-9)} (\tilde{A}_{IR, 0}-4 \tilde{A}_{IR, L})+\vr_0^{3/2} \tilde{V}_{IR, 0}\right)\right)+\\
    &-24(3-4 \vr_0)^2 \tilde{A}_{IR, L} \sqrt{-\vr_0^2 (10 \vr_0-9)} (10 \vr_0-3)+\\
    &-3 M^4 \sqrt{\vr_0} \tilde{V}_{IR, -1}+16 \left(100 \vr_0^3-220 \vr_0^2+159 \vr_0-36\right) \vr_0^{3/2} \tilde{V}_{IR, -1}+
    \\
    &2\log(\vr-\vr_0)\left(-4 M^2 \left(3 \tilde{A}_{IR, L} (4 \vr_0-3) \sqrt{-\vr_0^2 (10 \vr_0-9)}\right)+M^4 \sqrt{\vr_0} \tilde{V}_{IR, -1}\right)\Bigg]
    \\ &+\cdots \,,
    \end{split}
\end{equation}
where the four integration constants are denoted as $\tilde{A}_{IR, 0} $,  $\tilde{A}_{IR, L}$, $\tilde{V}_{IR, -1}$, and $\tilde{V}_{IR, 0}$.

We report also the IR expansion of the scalar fluctuations computed in the probe approximation, which are given by the following expressions:
\begin{equation}
    \begin{split}
    \mathfrak{p}^{\phi }_{IR}(\vr)=&\mathfrak{p}^{\phi }_{IR, 0} +\mathfrak{p}^{\phi }_{IR, L} \log(\vr-\vr_0)+ \\
    &\frac{(\vr-\vr_0)}{18-24 \vr_0}\Bigg[\frac{\mathfrak{p}^{\phi }_{IR, 0}\left(2 M^2-88 \vr_0^2+174 \vr_0-81\right)}{\vr_0}+\frac{2 \mathfrak{p}^{\phi}_{IR, L}\left(-2 M^2+128 \vr_0^2-204 \vr_0+81\right)}{\vr_0}\\
    &-\frac{M^2 \mathfrak{p}^{A_6}_{IR, 0}\sqrt{-\vr_0^2 (10 \vr_0-9)}}{\vr_0^{5/2}}+40 \mathfrak{p}^{\chi }_{IR, 0}\vr_0-\frac{27 \mathfrak{p}^{\chi}_{IR, 0}}{\vr_0}-6 \mathfrak{p}^{\chi}_{IR, 0}+40 \mathfrak{p}^{\chi}_{IR, L}\vr_0+\frac{54 \mathfrak{p}^{\chi}_{IR, L}}{\vr_0}\\
    &-96 \mathfrak{p}^{\chi}_{IR, L}-\frac{12 \mathfrak{p}^{A_6}_{IR, 1}\sqrt{-\vr_0^2 (10 \vr_0-9)}}{\sqrt{\vr_0}}+\frac{9 \mathfrak{p}^{A_6}_{IR, 1}\sqrt{-\vr_0^2 (10 \vr_0-9)}}{\vr_0^{3/2}}\Bigg]\\
    &+\frac{\frac{\mathfrak{p}^{\phi}_{IR, L}\left(2 M^2-128 \vr_0^2+210 \vr_0-81\right)}{\vr_0}+\frac{M^2 \mathfrak{p}^{A_6}_{IR, 0}\sqrt{-\vr_0^2 (10 \vr_0-9)}}{\vr_0^{5/2}}+\mathfrak{p}^{\chi}_{IR, L}\left(-80 \vr_0-\frac{27}{\vr_0}+102\right)}{18-24 \vr_0}\log(\vr-\vr_0)(\vr-\vr_0)
    \\ & +\cdots\,,
    \end{split}
\end{equation}
\begin{equation}
    \begin{split}
    \mathfrak{p}^{\chi }_{IR}(\vr)=&\mathfrak{p}^{\chi }_{IR, 0} +\mathfrak{p}^{\chi }_{IR, L} \log(\vr-\vr_0)- \\
    &\frac{(\vr-\vr_0)}{54-72 \vr_0}\Bigg[-\frac{6 M^2 \mathfrak{p}^{\chi}_{IR, 0}}{\vr_0}+\frac{12 M^2 \mathfrak{p}^{\chi}_{IR, L}}{\vr_0}+\frac{3 M^2 \mathfrak{p}^{A_6}_{IR, 0}\sqrt{-\vr_0^2 (10 \vr_0-9)}}{\vr_0^{5/2}}\\
    &+\mathfrak{p}^{\phi }_{IR, 0}\left(-40 \vr_0+\frac{27}{\vr_0}+6\right)+\mathfrak{p}^{\phi}_{IR, L}\left(-40 \vr_0-\frac{54}{\vr_0}+96\right)-200 \mathfrak{p}^{\chi }_{IR, 0}\vr_0\\
    &+\frac{9 \mathfrak{p}^{\chi}_{IR, 0}}{\vr_0}+162 \mathfrak{p}^{\chi}_{IR, 0}-80 \mathfrak{p}^{\chi}_{IR, L}\vr_0-\frac{18 \mathfrak{p}^{\chi}_{IR, L}}{\vr_0}+72 \mathfrak{p}^{\chi}_{IR, L}\\
    &+\frac{36 \mathfrak{p}^{A_6}_{IR, 1}\sqrt{-\vr_0^2 (10 \vr_0-9)}}{\sqrt{\vr_0}}-\frac{27 \mathfrak{p}^{A_6}_{IR, 1}\sqrt{-\vr_0^2 (10 \vr_0-9)}}{\vr_0^{3/2}}\Bigg] +\\
    &\frac{\frac{\mathfrak{p}^{\chi}_{IR, L}\left(6 M^2-160 \vr_0^2+162 \vr_0-9\right)}{\vr_0}+\frac{3 M^2 \mathfrak{p}^{A_6}_{IR, 0}\sqrt{-\vr_0^2 (10 \vr_0-9)}}{\vr_0^{5/2}}+\mathfrak{p}^{\phi}_{IR, L}\left(-80 \vr_0-\frac{27}{\vr_0}+102\right)}{54-72 \vr_0}\log(\vr-\vr_0)(\vr-\vr_0)
    \\ & +\cdots\,,
    \end{split}
\end{equation}
\begin{equation}
    \begin{split}
        \mathfrak{p}^{A_6 }_{IR}(\vr)=& \mathfrak{p}^{A_6 }_{IR, 0}+ \mathfrak{p}^{A_6 }_{IR, 1}(\vr-\vr_0)+\log(\vr-\vr_0)(\vr-\vr_0)\Bigg[-\frac{M^2 \mathfrak{p}^{A_6}_{IR, 0}+4 \sqrt{\vr_0} \sqrt{-\vr_0^2 (10 \vr_0-9)} (\mathfrak{p}^{\chi}_{IR, L}+3 \mathfrak{p}^{\chi}_{IR, L})}{3 \vr_0 (4 \vr_0-3)}\Bigg]
        \\ & +\cdots \,.
    \end{split}
\end{equation}
where we have denoted the six integration constants as $\mathfrak{p}^{\phi }_{IR, 0}$,  $\mathfrak{p}^{\phi }_{IR, L}$, $\mathfrak{p}^{\chi }_{IR, 0}$,  $\mathfrak{p}^{\chi }_{IR, L}$, $\mathfrak{p}^{A_6 }_{IR, 0}$, and $ \mathfrak{p}^{A_6 }_{IR, 1}$.

\subsection{UV expansions}
We collect  here the UV expansions, truncated to the first few terms, for all the spin-0 and spin-1 fluctuations of the soliton solutions with non-trivial flux. We find it convenient to write the expansion at asymptotically large values of $\vr$ as  powers of a fifth way to parametrise the holographic direction, by defining $\fz\equiv\frac{1}{\vr}$.
For the three gauge-invariant scalars we find the following expressions:
\begin{equation}
    \begin{split}
        &\mathfrak{a}^{\phi}_{UV}(\fz)=\mathfrak{a}^{\phi}_{UV,2} \fz^2+\mathfrak{a}^{\phi}_{UV,3} \fz^3-\frac{1}{8} \fz^4 \left(9 \mathfrak{a}^{\phi}_{UV,2} M^2\right)+\frac{1}{8} \fz^5 \left(-4 \mathfrak{a}^{\phi}_{UV,2} c^2-3 \mathfrak{a}^{\phi}_{UV,3} M^2\right)\\
        &+\frac{1}{128} \fz^6 \left(27 \mathfrak{a}^{\phi}_{UV,2} M^4+32 \mathfrak{a}^{\phi}_{UV,3} c^2\right)\\
        &+\frac{1}{640} (-9) \fz^7 \left(48 \mathfrak{a}^{\phi}_{UV,2} c^2 M^2-32 \mathfrak{a}^{\phi}_{UV,2} \mu -3 \mathfrak{a}^{\phi}_{UV,3} M^4-27 \sqrt{2} \mathfrak{a}^{A_6}_{UV,0} c \sqrt{\mu } M^2\right)+\cdots \,,
    \end{split}
\end{equation}
\begin{equation}
    \begin{split}
        &\ac_{UV}(\fz)=\ac_{UV, 0}+\frac{1}{8} \fz^2 \left(3 \mathfrak{a}^{\chi}_{UV,0} M^2\right)+\frac{1}{128} \fz^4 \left(27 \mathfrak{a}^{\chi}_{UV,0} M^4\right)\\
        &+\mathfrak{a}^{\chi}_{UV,5} \fz^5-\frac{\fz^6 \left(81 \mathfrak{a}^{\chi}_{UV,0} M^6\right)}{1024}\\
        &+\frac{1}{448} (-9) \fz^7 \left(-12 \mathfrak{a}^{\chi}_{UV,0} c^2 M^4-12 \mathfrak{a}^{\chi}_{UV,0} \mu  M^2+8 \mathfrak{a}^{\chi}_{UV,5} M^2-27 \sqrt{2} \mathfrak{a}^{A_6}_{UV,0} c \sqrt{\mu } M^2\right)+\cdots \,,
    \end{split}
\end{equation}
\begin{equation}
    \begin{split}
    &\mathfrak{a}^{A_6}_{UV}(\fz)= \mathfrak{a}^{A_6}_{UV, 0}+\frac{1}{8} \fz^2 \left(9 \mathfrak{a}^{A_6}_{UV,0} M^2\right)+\mathfrak{a}^{A_6}_{UV,3} \fz^3-\frac{1}{128} \fz^4 \left(81 \mathfrak{a}^{A_6}_{UV,0} M^4\right)\\
        &-\frac{3}{40} \fz^5 \left(8 \sqrt{2} \mathfrak{a}^{\phi}_{UV,2} c \sqrt{\mu }+9 \sqrt{2} \mathfrak{a}^{\chi}_{UV,0} c \sqrt{\mu } M^2-12 \mathfrak{a}^{A_6}_{UV,0} c^2 M^2+3 \mathfrak{a}^{A_6}_{UV,3} M^2\right)\\
        &+\frac{\fz^6 \left(-512 \sqrt{2} \mathfrak{a}^{\phi}_{UV,3} c \sqrt{\mu }+81 \mathfrak{a}^{A_6}_{UV,0} M^6+1024 \mathfrak{a}^{A_6}_{UV,3} c^2\right)}{1024}\\
        &+\frac{\fz^7 \left(27 \left(88 \sqrt{2} \mathfrak{a}^{\phi}_{UV,2} c \sqrt{\mu } M^2-36 \sqrt{2} \mathfrak{a}^{\chi}_{UV,0} c \sqrt{\mu } M^4-132 \mathfrak{a}^{A_6}_{UV,0} c^2 M^4-30 \mathfrak{a}^{A_6}_{UV,0} \mu  M^2+3 \mathfrak{a}^{A_6}_{UV,3} M^4\right)\right)}{4480}+\cdots\,.
    \end{split}
\end{equation}
The free parameters in these expansions are $\mathfrak{a}^{\phi}_{UV,2}$,  $\mathfrak{a}^{\phi}_{UV,3}$, $\ac_{UV, 0}$, $\mathfrak{a}^{\chi}_{UV,5}$, $\mathfrak{a}^{A_6}_{UV, 0}$, and $\mathfrak{a}^{A_6}_{UV,3}$, with the replacement $M^2=-q^2$,  while the other parameters, $c$ and $\mu$, are determined by the background.

For the to gauge-invariant (transverse) vector fluctuations, the  UV expansions can be written as follows:
 \begin{equation}
     \begin{split}
         \tilde{A}_{UV}(\fz)&=\tilde{A}_{UV, 0}+\frac{9 \tilde{A}_{UV, 0} M^2 \fz^2}{4\ 2}+\tilde{A}_{UV, 3} \fz^3-\frac{81}{128} M^4
   \tilde{A}_{UV, 0} \fz^4\\
        &+\frac{9}{80} \left(8 c^2 M^2 \tilde{A}_{uv, 0}-2 M^2 \tilde{A}_{UV, 3}-\sqrt{2}
   c M^2 \sqrt{\mu } \tilde{V}_{UV, 0}\right) \fz^5\\
        &+\frac{\left(81 M^6 \tilde{A}_{UV, 0}+1024 c^2
   \tilde{A}_{UV, 3}\right) \fz^6}{1024}\\
        &-\frac{\left(81 \left(44 c^2 M^4 \tilde{A}_{UV, 0}-40 M^2 \mu
    \tilde{A}_{UV, 0}-M^4 \tilde{A}_{uv, 3}+2 \sqrt{2} c M^4 \sqrt{\mu } \tilde{V}_{UV, 0}\right)\right)
   \fz^7}{4480} +\cdots\,,
     \end{split}
 \end{equation}
\begin{equation}
    \begin{split}
        \tilde{V}_{UV}(\fz)&=\tilde{V}_{UV, 0}+\tilde{V}_{uv, 0}+\frac{9 \tilde{V}_{UV, 0} M^2 \fz^2}{4\ 6}+\frac{27}{128} M^4 \tilde{V}_{UV, 0}
   \fz^4+\tilde{V}_{UV, 5} \fz^5\\
        &-\frac{\left(81 M^6 \tilde{V}_{UV, 0}\right) \fz^6}{1024}+\frac{9}{224}
   \left(54 \sqrt{2} c M^2 \sqrt{\mu } \tilde{A}_{UV, 0}+6 c^2 M^4 \tilde{V}_{UV, 0}+21 M^2 \mu 
   \tilde{V}_{UV, 0}-4 M^2 \tilde{V}_{UV, 5}\right) \fz^7+\cdots\,.
        \end{split}
\end{equation}
The free parameters are $\tilde{A}_{UV, 0}$, $\tilde{A}_{UV, 3}$, $\tilde{V}_{UV, 0}$, and $\tilde{V}_{UV, 5}$.

As for the gauge-invariant case, also in the case of the probe-approximation treatment of the  scalars, for each degree of freedom two modes appear in the UV expansion. The leading and sub-leading modes for $\{\mathfrak{p}^{\phi}_{UV}(\fz)$, $\mathfrak{p}^{\chi}_{UV}(\fz)$, and $\mathfrak{p}^{A_6}_{UV}(\fz)$ appear at the following orders:  $\{\fz^2,\fz^3\}, \{\fz^{\frac{1}{6} \left(-\sqrt{105}+15\right)} ,\fz^{\frac{1}{6} \left(\sqrt{105}+15\right)} \}$, and $\{1,\fz^3\}$, respectively. In view of the presence of  non-rational exponents, we find it convenient to show only the subleading terms, having imposed Dirichlet boundary conditions that remove the leading modes. This is the explicit form of the expansions used in our numerical analysis:
\begin{equation}
    \begin{split}
        &\mathfrak{p}^{\phi}_{UV}(\fz)=\mathfrak{p}^{\phi}_{UV,3} \fz^3-\frac{18}{48} M^2 \mathfrak{p}^{\phi}_{UV,3} \fz^5+\frac{\left(9 c^2 \mathfrak{p}^{\chi}_{UV,2}\right)
   \fz^{\frac{1}{6} \left(15+\sqrt{105}\right)+3}}{35+3 \sqrt{105}}+\frac{24}{96} c^2
   \mathfrak{p}^{\phi}_{UV,3} \fz^6+\cdots\,,
    \end{split}
\end{equation}
\begin{equation}
    \begin{split}
        &\mathfrak{p}^{\chi}_{UV}(\fz)=\mathfrak{p}^{\chi}_{UV,2} \fz^{\frac{1}{6} \left(\sqrt{105}+15\right)} \left(1-\frac{\left(27 M^2\right) \fz^2}{8 \left(\sqrt{105}+6\right)}+\frac{\left(\left(\sqrt{105}+10\right) c^2\right) \fz^3}{\sqrt{105}+9}\right)+\frac{3}{28} c^2 \mathfrak{p}^{\phi}_{UV,3} \fz^6+\cdots\,,
    \end{split}
\end{equation}
\begin{equation}
    \begin{split}
    \mathfrak{p}^{A_6}_{UV}(\fz)= &\mathfrak{p}^{A_6}_{UV,3} \fz^3-\frac{9}{40} M^2 \mathfrak{p}^{A_6}_{UV,3} \fz^5-\frac{\left(9
   \left(15+\sqrt{105}\right) c \mathfrak{p}^{\chi}_{UV,2} \sqrt{\mu }\right) \fz^{\frac{1}{6}
   \left(15+\sqrt{105}\right)+3}}{2 \sqrt{2} \left(25+2 \sqrt{105}\right)}
   \\ &+\left(c^2
   \mathfrak{p}^{A_6}_{UV,3}-\frac{c \mathfrak{p}^{\phi}_{UV,3} \sqrt{\mu }}{\sqrt{2}}\right) \fz^6+\cdots\,,
    \end{split}
\end{equation}
where the three non-vanishing integration constants are $\mathfrak{p}^{\phi}_{UV,3} $,  $\mathfrak{p}^{\chi}_{UV,2} $, and $\mathfrak{p}^{A_6}_{UV,3} $.

\bibliographystyle{JHEP} 
\bibliography{RC}

\end{document}